\documentclass[12pt]{article}

\usepackage{scicite}
\usepackage{amsmath}
\usepackage{amsfonts}
\usepackage{times}
\usepackage{graphicx}
\usepackage{textcomp}
\usepackage{float}
\usepackage{xcolor}
\usepackage{soul}
\usepackage{comment}
\usepackage{lineno}
\usepackage{amssymb}
\usepackage{verbatim}

\usepackage{color}

\usepackage[resetlabels]{multibib}
\newcites{supp}{Supplementary References}

\soulregister\cite{7} 
\soulregister\ref{7} 
\soulregister\eqref{7} 
\newenvironment{sciabstract}{
	\begin{quote} \bf}
	{\end{quote}}

\title{Anderson localization of electromagnetic waves in three dimensions}
\author
{Alexey Yamilov$^{1\star}$, Sergey E. Skipetrov$^{2}$, Tyler W. Hughes$^{3}$,  Momchil Minkov$^{3}$, \\ Zongfu Yu$^{3,4,\dagger}$,  Hui Cao$^{5\ast}$\\
	\\
	\normalsize{$^{1}$Physics Department, Missouri University of Science \& Technology, Rolla, Missouri 65409}\\
	\normalsize{$^{2}$Univ. Grenoble Alpes, CNRS, LPMMC, 38000 Grenoble, France}\\
	\normalsize{$^{3}$Flexcompute Inc, 130 Trapelo Road, Belmont, MA, 02478}\\
	\normalsize{$^{4}$Department of Electrical \& Computer Engineering, University of Wisconsin, Madison, WI, 53705}\\
	\normalsize{$^{5}$Department of Applied Physics, Yale University, New Haven, Connecticut 06520}\\
	\normalsize{$^\star$ yamilov@mst.edu}\\
	\normalsize{$^\dagger$ zongfu@flexcompute.com}\\ 
	\normalsize{$^\ast$ hui.cao@yale.edu}\\
}
\date{\today}

\begin{document} 
	\baselineskip24pt
	\maketitle 
	\begin{sciabstract}
		Anderson localization marks a halt of diffusive wave propagation in disordered systems.  Despite extensive studies over the past 40 years, Anderson localization of light in three dimensions has remained elusive, leading to the question of its very existence. Recent orders-of-magnitude speed-up of finite-difference time-domain calculations allows us to conduct brute-force numerical simulations of light transport in fully disordered 3D systems with unprecedented dimension and refractive index contrast. We demonstrate three-dimensional localization of vector electromagnetic waves in random packings of metallic spheres, in sharp contrast to the absence of localization for dielectric spheres with a refractive index contrast up to 10.  Our work opens a wide range of avenues in both fundamental research related to Anderson localization and potential applications using 3D localized light.
	\end{sciabstract}
	
	Anderson localization (AL) \cite{anderson58} is an emergent phenomenon for both quantum and classical waves including electron \cite{1967_Mott,1993_Kramer, 1998_Imada_Localization_review}, cold-atom \cite{2008_Aspect_Localization_of_Matter_Waves, jendr12}, electromagnetic (EM) \cite{john84,anderson85, chabanov00, schwartz07,segev2013anderson}, acoustic \cite{1985_Kirkpatrick, hu08}, water \cite{1983_Soillard_Water_Waves},  seismic \cite{1990_Sheng_Papanicolaou}, and gravity \cite{2013_Rothstein_Localization_Gravity} waves.
	Unlike in one or two dimensions, AL in three dimensions (3D) requires strong disorder  \cite{anderson58, john1991localization,2006_Sheng,2009_Lagendijk_PT}. A mobility edge separating the diffuse transport regime from AL can be estimated from the Ioffe-Regel criterion $k_{\text{eff}} \, \ell_s\sim 1$, where $k_{\text{eff}}$ is the effective wavenumber in the medium and $\ell_s$ is the scattering mean free path \cite{1960_Ioffe_criterion}. This criterion suggests two avenues to achieving localization: reduction of $k_{\text{eff}}$ or $\ell_s$. For EM waves, the former is realized by introducing partial order or spatial correlation in scatterer positions \cite{john84, 2020_Scheffold_localization}. In comparison, reaching localization of light in fully random photonic media by increasing scattering strength (decreasing $\ell_s$) turns out to be much more challenging \cite{beek12, sperling16}. Despite successful experiments in low-dimensional systems \cite{chabanov00, schwartz07, lahini2008anderson}, 3D localization remained stubbornly elusive \cite{2016_Skipetrov}, which triggered theoretical \cite{skipetrov14, 2021_Skipetrov_Longitudinal_modes} and experimental \cite{cobus22} studies of the mechanisms that impede it. 
	
	Anderson himself originally proposed ``{\it a system composed essentially of random waveguides near cut-off and random resonators, such as might be realized by a random packing of metallic balls of the right size}'' as ``{\it the ideal system}'' for localization of EM radiation \cite{anderson85}. 
	In practice, the absorption of metals obscures localization \cite{genack91, chabanov00}, and the experimental focus shifted to dielectric materials with low loss and high refractive index \cite{watson87, genack91, wiersma97, storzer06, sperling13}. However, even for dielectric systems, experimental artifacts due to residual absorption and inelastic scattering mar the data \cite{beek12, sperling16}. Numerically, these artifacts can be excluded, but 3D random systems of sufficiently large dimension and high refractive index contrast could not be simulated due to an extraordinarily long computational time required \cite{2009_Conti_3D_localization, pattelli18}. 
	
	Recent implementation of the finite-difference time-domain (FDTD) algorithm on emerging computing hardware has brought orders of magnitude speed-up of numerical calculation~\cite{2021_Flexcompute, 2021_Hughes_Flexcompute}. Using this highly-efficient hardware-optimized version of the FDTD method, we solve the Maxwell equations by brute force in 3D. It enables us to simulate sufficiently large systems and high refractive index contrast to address the following questions: can 3D AL of EM waves be achieved in fully random systems of dielectric scatterers? If not, can it occur in any other systems without aid of spatial correlations? 
	
	Answering these long-standing questions not only addresses the fundamental aspects of wave transport and localization across multiple disciplines, but also opens new research avenues and applications. For example, in topological photonics \cite{lu14}, the interplay between disorder and topological phenomena may be explored beyond the limit of weak disorder in low-dimensional systems \cite{stutzer18}. Also in cavity quantum electrodynamics with Anderson-localized modes \cite{sapienza10}, going 3D will avoid the out-of-plane loss inherent for 2D systems and cover the full angular range of propagation directions \cite{wiersma10}.  In addition to fundamental studies, disorder and scattering has been harnessed for various photonic device applications, but mostly with diffuse waves \cite{CaoAPR21}. Anderson-localized modes can be used for 3D energy confinement to enhance optical nonlinearities, light-matter interactions, and control random lasing as well as targeted energy deposition.    
	
\begin{figure}[H]
		\begin{center}
			\includegraphics[width=\textwidth]{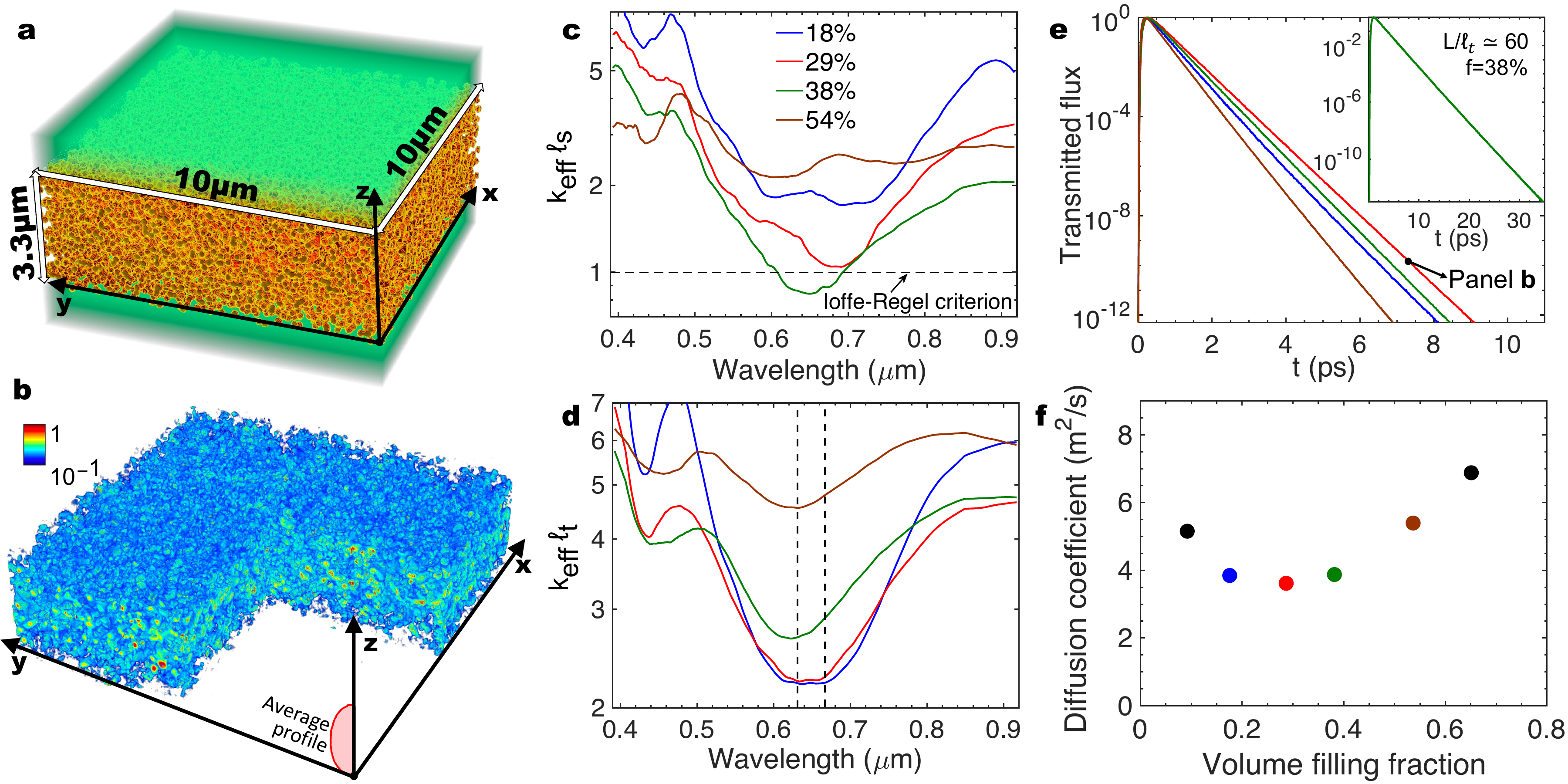}
			\caption{{\bf Absence of non-diffusive transport in random dielectric systems with  index contrast of 3.5.} 
				a,~3D slab filled with dielectric spheres at random uncorrelated positions (radius $r = 100$ nm, refractive index $n= 3.5$) in air. The slab cross-section is 10 $\mu$m$\times$10 $\mu$m = 100 $\mu$m$^2$ and thickness is $L = 3.3$ $\mu$m.  
				b,~3D distribution of light intensity inside the slab (dielectric filling fraction $f = 29$\%, $L/\ell_t = 33$) at long delay time after a short pulse of plane wave front is incident on the front surface. Red curve with shading shows the average depth profile.
				c,~Spectral dependence of the Ioffe-Regel parameter $k_{\text{eff}} \, \ell_s$ for different volume filling fractions of dielectric spheres, showing enhancement of scattering around single-sphere Mie resonances. The horizontal dashed line marks the Ioffe-Regel criterion $k_{\text{eff}} \, \ell_s =1 $  for 3D localization.   
				d,~Transport mean free path $\ell_t$ (in units of $1/k_{\text{eff}}$) as a function of wavelength, revealing a saturation by dependent scattering at high dielectric filling fractions. The vertical dashed lines mark spectral width (33 nm) of the excitation pulse in b and e.
				e,~Transmittance of the 3D slab for a pulsed excitation, showing exponential decay in time for all dielectric filling fractions, in agreement with diffusive transport. Inset shows persistence of diffusion when $L/\ell_t$ is increased from 33 to 60 for $f=38\%$, c.f. green line.  
				f,~Dependence of the minimum diffusion coefficient within the pulse bandwidth on the dielectric filling fraction $f$, exhibiting a minimum value of 3.6 m$^2$/s at $f\simeq 30\%$.
				\label{fig:n3.5}}
		\end{center}
	\end{figure}

	We first consider EM wave propagation through a 3D slab of randomly packed lossless dielectric spheres of radius $r=100$ nm and refractive index $n=3.5$. This corresponds to the highest index contrast achieved experimentally in optical range with porous GaP around the wavelength $\lambda=650$ nm in the vicinity of the first Mie resonance of an isolated sphere, see Supplementary Information (SI). To avoid spatial correlations, the sphere positions are chosen completely randomly, leading to spatial overlap where index is capped at the same value of $n$. We compute the spatial correlation function of such structure, which reveals that the correlation vanishes beyond the particle diameter (SI). To avoid light reflection at the interfaces of the slab, we surround it by a uniform medium with the refractive index equal to the effective index of the slab,  $n_{\text{eff}}=[(1-f)+f n^2]^{1/2}$, for a given dielectric volume filling fraction $f$ (Fig.~\ref{fig:n3.5}a). As described in SI, for each wavelength, we compute the scattering mean free path $\ell_s$ directly from the rate of attenuation of co-polarized field with depth. This, together with the effective wavenumber $k_{\text{eff}}=n_{\text{eff}} \, (2\pi/\lambda)$, gives the Ioffe-Regel parameter plotted in Fig.\ \ref{fig:n3.5}c. It features a minimum around $\lambda = 650$ nm, and the smallest value of $k_{\text{eff}} \, \ell_s \simeq 0.9$ is reached at $f=38\%$. We also compute the transport mean free path $\ell_t$ from continuous wave (CW) transmittance of an optically thick slab with thickness $L \gg \ell_t$  (SI). In Fig.\ \ref{fig:n3.5}d, $k_{\text{eff}} \, \ell_t$ also exhibits a dip in the same wavelength range as $k_{\text{eff}} \, \ell_s$, however, the smallest $k_{\text{eff}} \, \ell_t$ is found at lower $f$ of $18$--29\%, as the dependent scattering sets in at higher $f$.
	In search for AL in this wavelength range, we numerically simulate propagation of a narrowband Gaussian pulse centered at $\lambda_0 = 650$ nm with plane wavefront, and compute the transmittance through the slab $T(t)$ as a function of arrival time $t$. The diffusive propagation time $\tau_D$ approximately corresponds to the arrival time of the peak in Fig. \ref{fig:n3.5}e. At $t \gg \tau_D$, the decay of the transmitted flux is exponential over $12$ orders of magnitude, as expected for purely diffusive systems \cite{2007_Akkermans_book}. Rate of this exponential decay is equal to $1/\tau_D$, which is directly related\cite{2007_Akkermans_book} to the smallest diffusion coefficient within the spectral range of the excitation pulse (SI). In Fig. \ref{fig:n3.5}f, the dependence of this diffusion coefficient $D$ on the dielectric filling fraction $f$ exhibits a minimum at $f\sim 30\%$. Inset in Fig. \ref{fig:n3.5}e shows that the further increase of the slab thickness does not lead to any deviation from diffusive transport. Furthermore, the diffusive behavior persists in the numerical simulation with increased spatio-temporal resolution (SI). At $t \gg \tau_D$, the spatial intensity distribution inside the system features a depth profile (averaged over cross-section) equal to that of the first eigenmode of the diffusion equation  (Fig. \ref{fig:n3.5}b). We therefore rule out a possibility of AL in uncorrelated ensembles of dielectric spheres with $n=3.5$.
	
	At microwave frequencies, the refractive index may be even higher than $n=3.5$. We therefore, perform numerical simulation of a 3D slab of dielectric spheres with $n=10$. The main results are summarized here, and details are presented in SI. A large scattering cross section $\sigma_s(\lambda)$ of a single sphere near the first Mie resonance leads to strong dependent scattering already at small filling fractions. We find the Ioffe-Regel parameter $k_{\text{eff}} \, \ell_s \gtrsim 1$ despite the very high refractive index contrast. This is attributed to dependent scattering that becomes significant even at relatively low dielectric filling fraction $f$. The numerically calculated $T(t)$  for $L/\ell_t \gg1$ does not exhibit any deviation from diffusive transport: at $t\gg \tau_D$, the decay of transmittance is still exponential over $\sim 10$ orders of magnitude. In addition, scaling of CW transmittance with the inverse slab thickness $1/L$ remains linear for all $f$, as expected for diffusion (SI). We therefore conclude that AL does not occur in random ensembles of dielectric spheres, thus closing the debate about the possibility of light localization in a white paint \cite{anderson85, sperling16}. 
	
	Previous studies \cite{skipetrov14, 2021_Skipetrov_Longitudinal_modes} suggest that absence of AL for EM waves may be due to longitudinal waves that exist in a heterogeneous dielectric medium, where the transversality condition $\nabla \cdot \mathbf{E}(\mathbf{r}) = 0$ does not follow from the Gauss's law $\nabla \cdot [\epsilon(\mathbf{r}) \mathbf{E}(\mathbf{r})] = 0$, because of the position dependence of $\epsilon(\mathbf{r})$. Here, we propose to suppress the contribution of longitudinal waves to optical transport and realize AL of EM waves by using perfectly conducting spheres as scatterers. The Poynting vector is parallel to the surface of a perfect electric conductor (PEC) \cite{1981_van_de_Hulst} and EM energy flows around a PEC particle without coupling to longitudinal surface modes. The volume of PEC spheres is simply excluded from the free space and becomes unavailable for light. Thus, at high PEC volume fraction, light propagates in a random network of irregular air cavities and waveguides formed by the overlapping PEC spheres, akin to the original proposal of Anderson \cite{anderson85}.
	
\begin{figure}[H]
		\begin{center}
			\includegraphics[width=\textwidth]{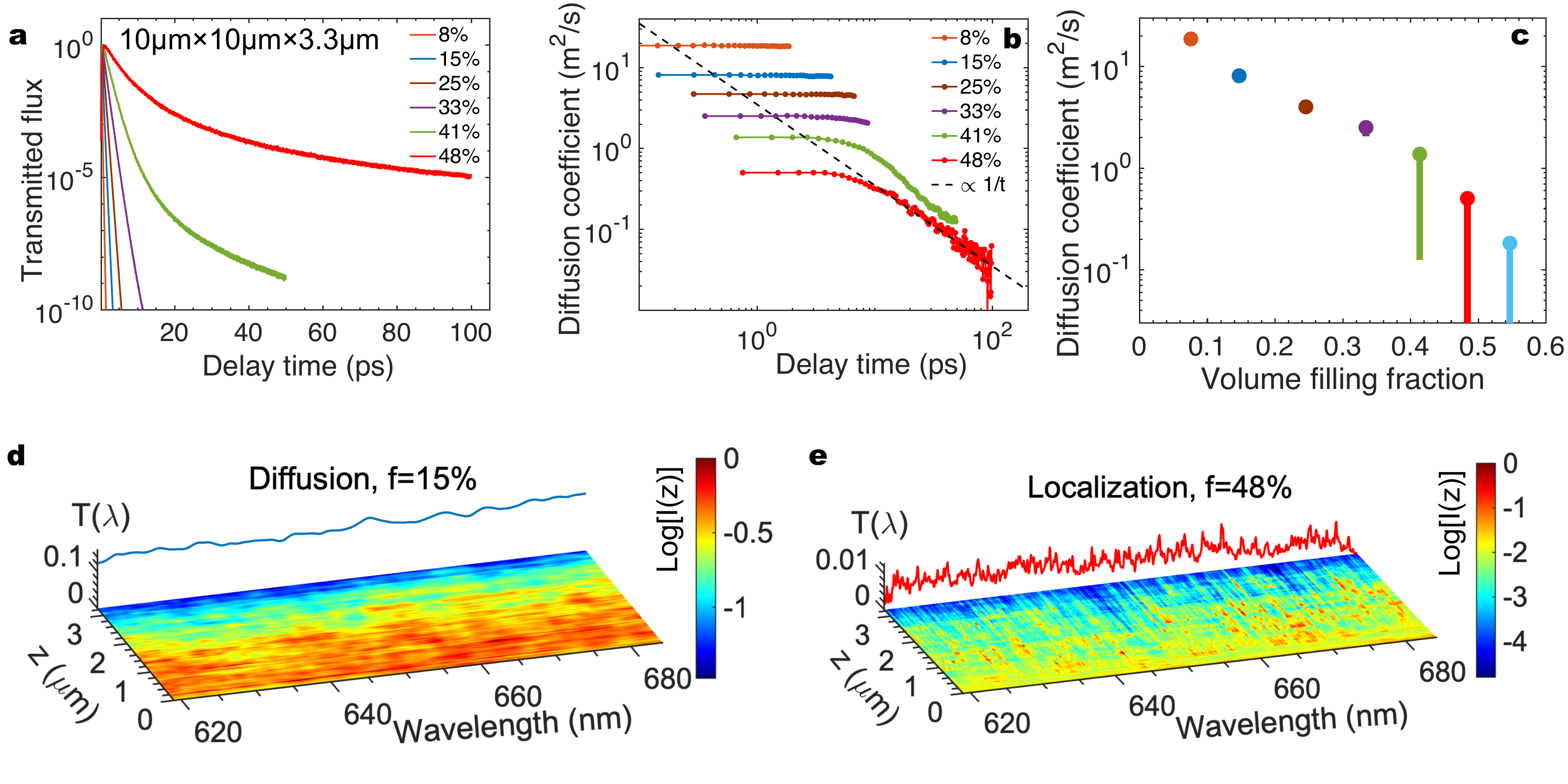}
			\caption{{\bf Anderson localization of light in 3D disordered PEC.} 
				a,~Transmittance $T(t)$ of an optical pulse through a 3D slab (10 $\mu$m $\times$ 10 $\mu$m $\times$ 3.3 $\mu$m) of randomly packed PEC spheres with radius $r=50$ nm and volume filling fractions $f$ from 8\% to 48\%. 
				b,~Time-resolved diffusion coefficient $D(t)$ extracted from the decay rate of $T(t)$ in panel a, decreasing with time as $1/t$ at high $f$. 
				c,~Short-time $D$ (dots) and the interval of variation of $D$ with time (bars) at different $f$.  
				d,e,~Continuous-wave transmittance spectrum $T(\lambda)$ in diffusive (d, $f= 15\%$, blue line) and localized (e, $f= 48\%$, red line) PEC slabs. Color map: depth profile of average intensity $\langle I(x,y_0,z;\lambda)\rangle_x$ inside the slab at different wavelengths, highlighting the localized and necklace-like states for $f = 48$\%. 
				\label{fig:PEC}}
		\end{center}
	\end{figure}

	Similarly to the dielectric systems above, we simulate a 3D slab composed of randomly packed, overlapping PEC spheres of radius $r=50$ nm in air. Figure~\ref{fig:PEC} shows the results of simulating an optical pulse propagating through 10 $\mu$m$\times$10 $\mu$m$\times$3.3 $\mu$m slabs of various PEC fractions. $T(t)$ displays non-exponential tails at high fractions $f=41\%,\ 48\%$ in Fig. \ref{fig:PEC}a. From the decay rate obtained via a sliding-window fit, we extract a time-dependent diffusion coefficient $D(t)$, c.f. Fig. \ref{fig:PEC}b, which shows a power-law decay with time, as predicted by the self-consistent theory of localization\cite{2006_Skipetrov_dynamics}. The non-exponential decay of $T(t)$ and the time-dependence of $D$ are the signatures of AL \cite{2006_Skipetrov_dynamics,hu08}. In contrast, at lower PEC fractions $f=8\%,\ 15\%$, $D$ remains constant in time. Figure~\ref{fig:PEC}c reveals a transition from time-invariant $D$ to time-dependent $D(t)$ around $f=33\%$ where $D(t)$ starts deviating from a constant. Using Fourier transform, we compute the spectrally resolved transmittance $T(\lambda)$. Figures~\ref{fig:PEC}d,e contrast $T(\lambda)$ of diffusive and localized systems. The former features smooth, gradual variations with $\lambda$ due to broad overlapping resonances, whereas the latter exhibits strong resonant structures consistent with the average mode spacing exceeding the linewidth of individual modes, in accordance with Thouless criterion of localization, as the spectral narrowing of modes is intimately related to their spatial confinement \cite{1974_Thouless, 1993_Kramer, chabanov00}. Color maps in Figs.~\ref{fig:PEC}d,e show spatial intensity distributions inside the systems, $\langle I(x,y_0,z;\lambda)\rangle_x$, averaged over $x$ for a cross-section $y=y_0$. These two-dimensional maps contrast slow variation with $z$ and $\lambda$ in the diffusive system (panel d) to the sharp features due to spatially-confined modes in the localized system (panel e). Furthermore, there exist `necklace' states with multiple spatially separated intensity maxima, originally predicted for electrons in metals \cite{1987_Pendry}.
	
\begin{figure}[H]
		\begin{center}
			\includegraphics[width=\textwidth]{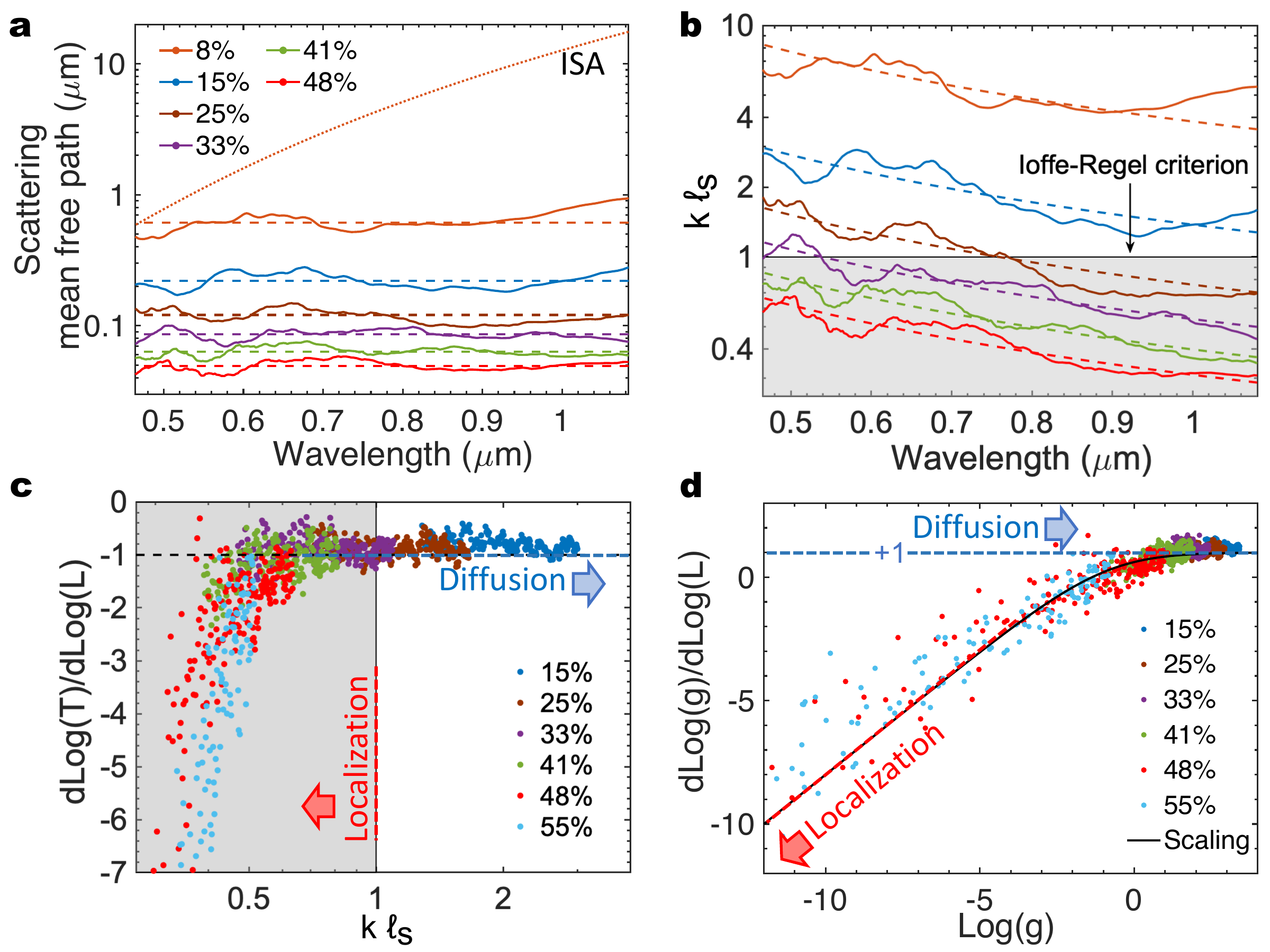}
			\caption{{\bf Transition from diffusion to Anderson localization in 3D disordered PEC.} a,~Scattering mean free path $\ell_s$ for PEC volume filling fractions $f$ from 8\% to 48\%. $\ell_s$ is nearly flat over broad spectral range. 
				b,~Spectral dependence of Ioffe-Regel parameter $k \ell_s$, exhibiting $1/\lambda$ dependence (dashed lines). 
				c,~Scaling of the continuous-wave (CW) transmittance $T$ with slab thickness $L$ versus Ioffe-Regel parameter $k \ell_s$, revealing diffusion-localization transition at $k \ell_s\sim 1$. Blue dashed line denotes diffusive scaling $T\propto L^{-1}$; red dashed line marks $k \ell_s = 1$.
				d,~Single-parameter scaling of dimensionless conductance $g$ for diffusion-localization transition (solid black line), in agreement with numerical data for 6 PEC filling fractions. Blue and red dashed lines denote diffusive and localized scalings $g\propto L$ and $g \propto \exp(-L/\xi)$ respectively, where $\xi$ is the localization length.
				\label{fig:PEC_scaling}}
		\end{center}
	\end{figure}

	Insight into the mechanism behind AL in the random ensemble of PEC spheres can be gained from the wavelength dependence of the Ioffe-Regel parameter $k \ell_s$. We compute it using a procedure similar to that in dielectrics (SI). Even at the volume fraction of $f = 8\%$, $\ell_s$ is well below the prediction of independent scattering approximation (ISA), due to scattering resonances formed by two or more neighboring PEC spheres (SI). As shown in Fig.~\ref{fig:PEC_scaling}a, $\ell_s$ becomes essentially independent of wavelength in the range of size parameter $k r$ of PEC spheres. Consequently, the Ioffe-Regel parameter acquires $1/\lambda$ dependence as seen in Fig.~\ref{fig:PEC_scaling}b. It drops below the value of unity within the excitation pulse bandwidth  $\lambda\simeq 650$ nm $\pm$ 45 nm for $f$ between $25\%$ and $33\%$, in agreement with Fig.~\ref{fig:PEC}. We further conduct a finite-size scaling study, after computing the dependence of CW transmittance $T$ on the slab thickness $L$ (SI). Figure~\ref{fig:PEC_scaling}c shows the logarithmic derivative $d\log(T)/d\log(L)$ as a function of $k \ell_s$. In the diffusive regime, Ohm's law $T\propto 1/L$ is expected, leading to a scaling power of $-1$, as indeed observed for $k \ell_s>1$. Around $k \ell_s\sim 1$ we see a departure from $1/L$ scaling of transmittance. Note that the scaling theory of localization\cite{abrahams79} predicts a single-parameter scaling with the dimensionless conductance $g$ but not with $k \, \ell_s$. By estimating the number of transverse modes as $N = 2\pi (L/\lambda)^2(1-f)^{2/3}$ for $L\times L$ area of the slab, we compute $g=TN$\cite{2011_Delande_lecture_notes} and $\beta(g)\equiv d\log(g)/d\log(L)$. Figure~\ref{fig:PEC_scaling}d shows a good agreement between the numerical data and the model function $\beta(g)=2-(1+g)\log(1+g^{-1})$\cite{2011_Delande_lecture_notes}. In diffusive regime $g>1$, $\beta(g)\rightarrow 1$; whereas in the localized regime $g<1$, $\beta(g)\propto\log(g)$. The latter is a manifestation of the negative exponential scaling of $g$ with $L$ in the regime of Anderson localization. 
	
\begin{figure}[H]
		\begin{center}
			\includegraphics[width=4in]{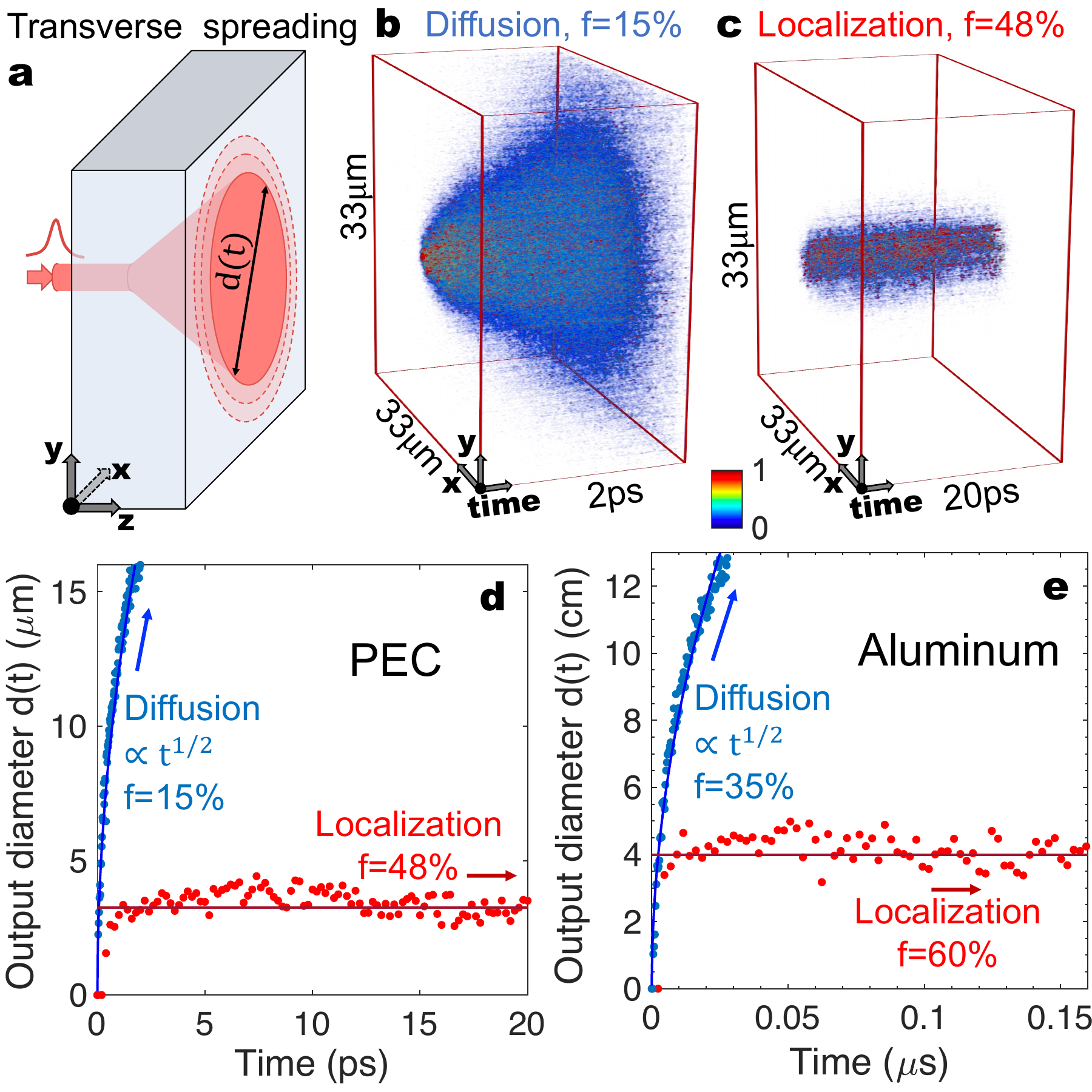}
			\caption{{\bf Arrest of transverse spreading of transmitted beam in 3D localized PEC systems.} a,~Schematic of transverse spreading of a tightly-focused pulse propagating through a diffusive slab of cross section 33 $\mu$m$\times$33 $\mu$m and thickness $L=3.3$  $\mu$m. b,~2D intensity distribution at the output surface (normalized to the maximum) for different delay times, showing a lateral expansion of the beam in the diffusive slab of PEC filling fraction $f=15\%$. c,~Absence of transverse spreading for $f=48\%$, due to Anderson localization. d,~Lateral diameter of the transmitted beam $d(t)$ increases as $\sqrt{t}$ in the diffusive slab (blue dots), but it saturates to a constant value in the localized slab (red dots). e,~Same as d but for a slab ($L = 6$ cm) of aluminum spheres ($r=0.28$ cm, $f=35\%$ and $60\%$) with realistic conductivity $\sigma_0=3.8\times 10^{4}$ ${\rm \Omega^{-1}/m}$ at microwave frequency $\sim 20$ GHz.
			\label{fig:PEC_spreading}}
		\end{center}
	\end{figure}

	To obtain the ultimate confirmation of AL of light in PEC composites, we simulate dynamics of the transverse spreading of a tightly focused pulse---a measurement that has been widely adopted in localization experiments\cite{2010_van_Tiggelen, hu08, sperling13}. A pulse centered at $\lambda = 650$ nm with a bandwidth of 90 nm is focused to a small spot of area $\simeq 0.5$ $\mu \rm{m}^2$ at the front surface of a wide 3D slab of dimensions 33 $\mu$m$\times$33 $\mu$m$\times$3.3 $\mu$m, c.f. Fig.~\ref{fig:PEC_spreading}a. We compute the transverse extent of intensity distribution $I(x,y,z=L;t)$ at the back surface of the slab. For a diffusive PEC slab with $f=15\%$, we observe a rapid transverse spreading of light with time in Fig.~\ref{fig:PEC_spreading}b, which approaches the lateral boundary of the slab within $\sim$ 2 ps. In sharp contrast, in the localized system in Fig.~\ref{fig:PEC_spreading}c ($f = 48$\%), the transmitted intensity profile remains transversely confined even after 20 ps. This time corresponds to a free space propagation of 6 mm, which is $\sim 2000$ times longer than the actual thickness of the slab. Figure~\ref{fig:PEC_spreading}d quantifies this time evolution with the output beam diameter $d(t)=2[\text{PR}(t)/\pi]^{1/2}$, where $\text{PR}(t)=[\int\int I(x,y,L;t) dx dy]^2/\int\int I(x,y,L;t)^2 dx dy$ is the intensity participation ratio. For a diffusive slab, $d(t) \propto t^{1/2}$, while in the localized regime, $d(t)$ saturates at a value of the order of slab thickness $L$. Such an arrest of the transverse spreading in the localized PEC systems persists with increased spatio-temporal resolution of the numerical simulation (SI). Further evidence of AL include non-linear depth profile and strong non-Gaussian fluctuations of intensity inside the system  (SI). We also confirm our results by repeating calculations for 3D slabs of PEC spheres with larger radius $r = 100$ nm, and obtaining similar scaling behavior (SI), as in Figs.~\ref{fig:PEC_scaling}cd.
	
	The striking difference between light propagation in dense random ensembles of dielectric and PEC spheres cannot be accounted for by the Ioffe-Regel parameter, as both reach $k_{\text{eff}} \, \ell_s\sim 1$ for similar values of the size parameter $kr$ (Figs.~\ref{fig:n3.5}c, \ref{fig:PEC_scaling}b). AL in 3D PEC composites with uncorrelated disorder reveals a localization mechanism that is unique for PEC. In contrast to a dielectric system where light propagates everywhere (both inside and outside the scatterers), the propagation is restricted to the voids between scatterers in the PEC system. AL takes place when the typical size of voids becomes smaller than the wavelength, and light can no longer ‘squeeze’ through them. This qualitative model correctly predicts both AL in the long-wavelength regime and the increase of the critical volume fraction $f$ for localization with the scatterer size $r$  (SI).
	
	Finally, we test AL in real-metal aggregates. In the microwave spectral region, the skin depth of crystalline metals like silver, aluminum and copper is several orders of magnitude shorter than the wavelength $\lambda$ and the scatterer size $r$ in the regime of $k r \sim 1$. Since the microwave barely penetrates into the metallic scatterers, our simulation results are almost identical to those for PEC. To account for the imperfections due to polycrystallinity, surface defects, oxide layers etc., we lower the metal conductivity to match the experimentally measured absorption rate in aggregates of aluminum spheres \cite{genack91}. Simulations unambiguously show the arrest of transverse spreading of a focused pulse (Fig.~\ref{fig:PEC_spreading}e), demonstrating AL in 3D random aggregates of aluminum spheres. Additional evidence of AL is presented in SI. Moreover, even at optical frequencies, where realistic metals deviate notably from PEC, the arrest of transverse spreading persists in 3D silver nanocomposites, as shown in SI. Possible light localization in 3D nanoporous metals will have a profound impact on their applications in photo-catalysis, optical sensing, energy conversion and storage.	
	
	In summary, our large-scale microscopic simulations of EM wave propagation in 3D uncorrelated random ensembles of spheres show no signs of Anderson localization for dielectric spheres with refractive indices $n=3.5$--10. This may explain multiple failed attempts of experimental observation of AL of light in 3D dielectric systems over the last three decades \cite{wiersma97, storzer06, sperling13, beek12,sperling16}. At the same time, we report the first numerical evidence of EM wave localization in random ensembles of PEC spheres over a broad spectral range. Localization is confirmed by eight criteria: the Ioffe-Regel criterion, the Thouless criterion, non-exponential decay of transmittance under pulsed excitation, vanishing of the diffusion coefficient, existence of spatially localized states, scaling of conductance, arrest of the transverse spreading of a narrow beam, and enhanced non-Gaussian fluctuations of intensity. Our study calls for renewed experimental efforts to be directed at low-loss metallic random systems \cite{genack91}. In the SI, we propose a realistic microwave experiment that avoids experimental pitfalls~\cite{chabanov00} and provides a tell-tale sign of Anderson localization.
	
	\bibliographystyle{Nature}
    \bibliography{loc3d_Nature_Physics.bbl}	
	
	\newpage
	\newpage
	\newpage
	\newpage
	\newpage
	\section*{Acknowledgments}
	This work is supported by the National Science Foundation under Grant Nos. DMR-1905465, DMR-1905442, and the Office of Naval Research (ONR) under Grant No. N00014-20-1-2197. We thank Shanhui Fan and Bart van Tiggelen for enlightening discussions.
	\section*{Conflict of interest statement}
    TWH, MM,  ZY have financial interest in Flexcompute Inc., which develops the software Tidy3D used in this work.
	\section*{Author contributions}
	A.Y. performed numerical simulations, analyzed the data and compiled all results. S.E.S. conducted theoretical study and guided data interpretation. T.W.H. and M.M. implemented the hardware-accelerated FDTD method and aided in the setup of the numerical simulations. Z.Y. and H.C. initiated this project and supervised the research.
	A.Y. wrote the first draft, S.E.S. and H.C. revised the content and scope, T.W.H. and M.M. and Z. Y. edited the manuscript.  All co-authors discussed and approved the content.
	\newpage
	\setcounter{page}{1}
	\begin{center}\section*{Supplementary Information:\\ 
			Anderson localization of electromagnetic waves in three dimensions}
		\author
		{Alexey Yamilov$^{1\star}$, Sergey E. Skipetrov$^{2}$, Tyler W. Hughes$^{3}$,  Momchil Minkov$^{3}$, \\ Zongfu Yu$^{3,4,\dagger}$,  Hui Cao$^{5\ast}$\\ \ \\
			\normalsize{$^{1}$Physics Department, Missouri University of Science \& Technology,Rolla, Missouri 65409}\\
			\normalsize{$^{2}$Univ. Grenoble Alpes, CNRS, LPMMC, 38000 Grenoble, France}\\
			\normalsize{$^{3}$Flexcompute Inc, 130 Trapelo Road, Belmont, MA, 02478}\\
			\normalsize{$^{4}$Dept. of Electrical \& Computer Engineering, University of Wisconsin, Madison, WI, 53705}\\
			\normalsize{$^{5}$Department of Applied Physics, Yale University, New Haven, Connecticut 06520}\\
			\normalsize{$^\star$ yamilov@mst.edu}\\
			\normalsize{$^\dagger$ zongfu@flexcompute.com}\\ 
			\normalsize{$^\ast$ hui.cao@yale.edu}\\
		}
		\maketitle
		\ \\
	\end{center}
	\renewcommand{\thefigure}{S\arabic{figure}}
	\setcounter{figure}{0}

	\makeatletter
	\renewcommand\@bibitem[1]{\item\if@filesw \immediate\write\@auxout
		{\string\bibcite{#1}{S\the\value{\@listctr}}}\fi\ignorespaces}
	\def\@biblabel#1{[S#1]}
	\makeatother

	\section{Methods}
	
	\subsection{Numerical method}
	
	Over the years, various techniques have been developed for large-scale 3D computations. One efficient method for simulating light scattering in large inhomogeneous media is based on Born series \citesupp{osnabrugge2016convergent, kruger2017solution}. This method, however, is limited to small contrast in refractive index between the scatterers and the host medium. To achieve 3D localization, a large refractive-index contrast is needed to enhance light scattering, which makes it necessary to include many terms of the Born series. This greatly increases the computational complexity, worsening the convergence of such an iterative algorithm.
	
	Another family of approaches for simulating wave transport in 3D aggregates of spherical particles is based on iterative multi-sphere scattering expansions, e.g., the generalized multi-particle-Mie (GMM) formalism~\citesupp{2001_Xu_GMM, 2007_Pellegrini_GMM, 2022_Egel_GMM}, the multi-sphere T-matrix (MSTM)~\citesupp{2011_Mackowski_MSTM}, and the fast multipole method (FMM)~\citesupp{gumerov2005computation}. The latest, and the most advanced implementation, deployed on GPUs for speedup, is capable of simulating up to $10^5$ dielectric spheres~\citesupp{2017_Wiersma_3D_simulations},\cite{pattelli18}.
	But the state-of-the-art algorithm (CELES) has not been used for index contrast higher than $2$. This is because the Mie resonances of high-index spheres can have very long lifetimes and strong internal fields, leading to an extremely slow convergence of the iterative algorithm or even its failure. Moreover, the spheres cannot touch or overlap, which would invalidate the Mie solution for isolated spheres. If two spheres are very close to each other, the near-field effects will excite high-order multipoles, causing slow or failed convergence. Therefore, these methods require a minimum distance between densely-packed spheres, which introduces spatial correlations of the scattering structures~\cite{pattelli18} that are known to affect Anderson localization~\citesupp{2008_Conti_AL_laser},\cite{2020_Scheffold_localization}.
	
	The finite-difference time-domain (FDTD) method directly solves the Maxwell's equations in space and time~\citesupp{2005_Taflove}. It does not make any physical approximations, and is capable of simulating large-refractive-index, spatially-overlapping or touching particles with high filling fraction. The effects of spatial correlations of scattering structures that could contribute to Anderson localization~\citesupp{2008_Conti_AL_laser},\cite{2020_Scheffold_localization}, are avoided by completely-random placement of individual scatterers in our study, as shown in Sec.~\ref{suppsec:spatial_corr} below. 
	
	The ability of simulating refractive-index contrast of 3.5--10 in our study is the key to demonstrate that increasing index contrast will not precipitate Anderson localization. However, simulating such high refractive-index particles requires very fine spatial and temporal resolution, leading to an extremely long computational time. Previous FDTD simulations were limited to small 3D structures containing $10^3$--$10^4$ particles \cite{2009_Conti_3D_localization},\cite{ 2020_Scheffold_localization}. Our hardware-accelerated implementation of the FDTD method~\cite{2021_Flexcompute} reduces the computational time by several orders of magnitude, allowing us to simulate a 3D system with $6 \times 10^6$ scatterers in about 40 min.
	
	The orders-of-magnitude reduction in computational time is  essential to {\it search} for Anderson localization in 3D PEC and real-metal composites, because it allows to repeat the simulations many times with varying parameters like particle size, volume fraction, dielectric function and conductivity, lateral dimension and thickness of simulated systems, etc. Moreover, our time-domain simulation can extract the fields at more than $1000$ discrete frequencies from a single run, by Fourier transform. In contrast, the frequency-domain methods based on Born series and GMM/MSTM simulate only one frequency per run. Using such methods to simulate time-dependent transport, particularly at long delay time, requires repeating the calculations at many closely-spaced frequencies so that the Fourier transform will give the long-time behavior.
	
	\subsection{System geometry and dimension\label{suppsec:geometry}}
	Unless otherwise specified, our simulations are carried out on disordered systems in the slab geometry $L_x=L_y\gg~L$ (Fig.~\ref{fig:n3.5}a). A plane wave with the electric field polarized along $x$-axis is incident on the front surface of the slab.  The bandwidth of a Gaussian pulse, $\Delta \lambda =$ 90 nm, is chosen such that $\ell_t$ remains nearly invariant with wavelength. Periodic boundary conditions are applied along $x$ and $y$ axes. To ensure that the periodicity would not affect the results presented in this manuscript, we make the transverse dimensions $L_x, L_y$ of the simulated slabs much larger than the thickness $L$. In the transmission simulations such as those in Fig.~\ref{fig:n3.5}, the ratio is $L_x/L=L_y/L=3$, so that any effect induced by periodic boundary conditions occurs on a length scale much larger than any relevant transport or localization scales. For the simulations of the transverse spreading (Figs.~\ref{fig:PEC_spreading},\ref{suppfig:Discretization_PEC},\ref{suppfig:optical_spreading}), we simulate much wider slabs having $L_x/L=L_y/L=10$ to avoid any edge effect.
	
	The slab is sandwiched between homogeneous layers of refractive index $n_B$ and thickness $\Delta_0$, which in turn are surrounded by perfectly matched layers of thickness $\Delta_{PML}$. These thicknesses are chosen as $\Delta_B=\Delta_{PML}=2\lambda_0,\lambda_0,2\lambda_0$ for the slabs of dielectric spheres with refractive index $n=3.5,10$ and PEC spheres, respectively. For the dielectric slabs, $n_0$ is equal to the effective index $n_{\text{eff}}$ found by averaging the dielectric constant, while for PEC $n_{0} = 1$. $\lambda_0=650$ nm is the central wavelength of optical pulses in all simulations. The spatial discretization step of the FDTD algorithm is $\lambda_0/20$ for $n=3.5$ and PEC, and $\lambda_0/60$ for $n=10$. This corresponded to time steps of $dt\simeq56\times10^{-18}$ s and $dt\simeq19\times10^{-18}$ s, respectively. 
	
	\subsection{Numerical accuracy and scaling\label{suppsec:accuracy}}
	Tidy3D solver\cite{2021_Flexcompute} is an implementation of the standard FDTD method, which does not make any physical approximation or impose any constraint of the scattering structures. The remaining numerical issue in FDTD simulation is sufficient discretization of space and time~\citesupp{2005_Taflove}. We have carefully tested spatio-temporal resolution of FDTD simulations to ensure consistency of our numerical results so that the conclusions of our study are robust and independent of the discretization. We present two examples of such tests below.
	 				
\begin{figure}[H]
		\begin{center}
			\includegraphics[width=5.2in]{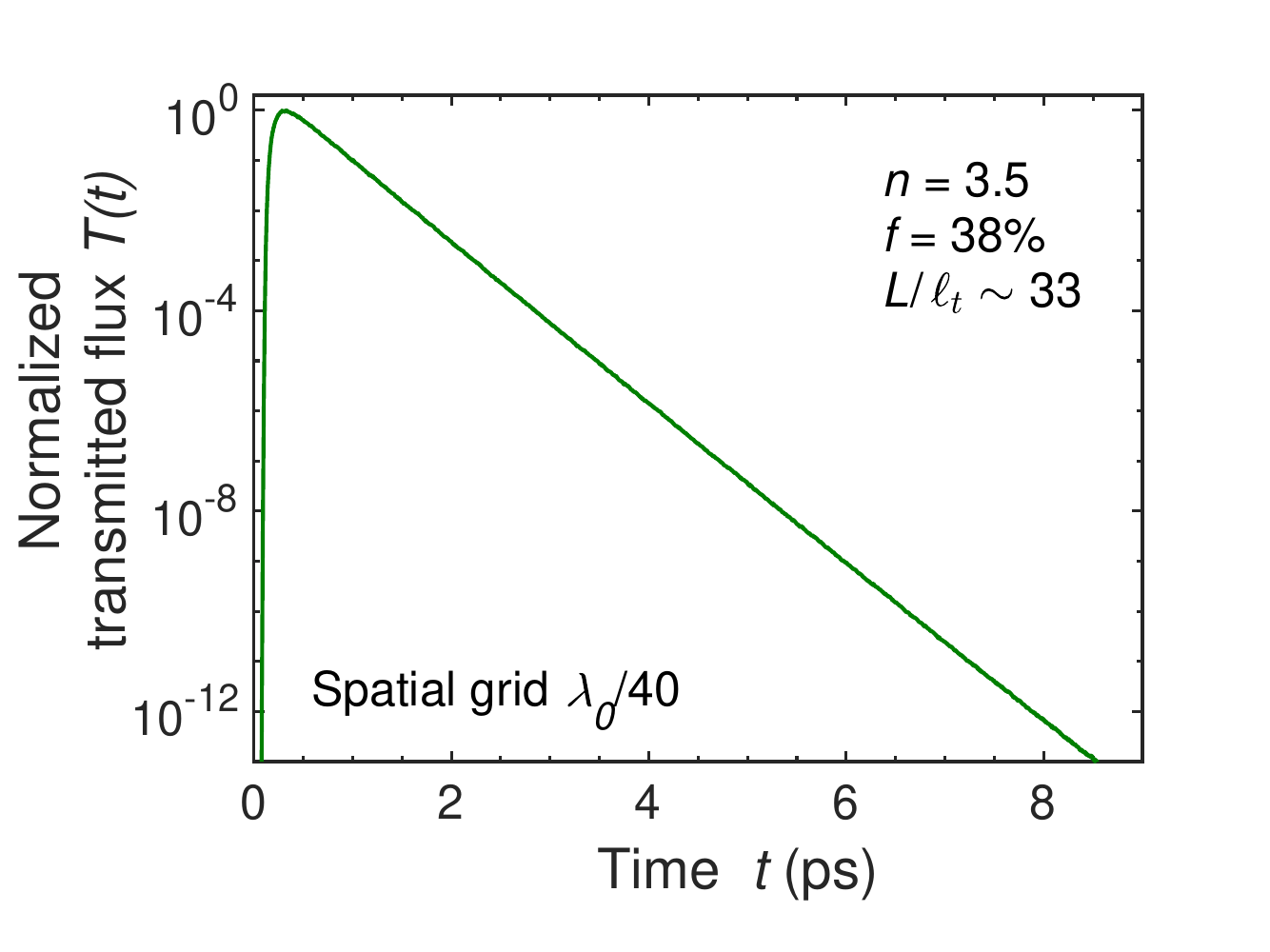}
			\caption{{\bf Confirmation of the absence of Anderson localization in random dielectric media with refractive-index contrast of $3.5$ with increased numerical resolution.} 
			Transmittance $T(t)$ of an optical pulse through a 3D slab of thickness $L = 3.3$ $\mu$m, filled with dielectric spheres at random uncorrelated positions (radius $r = 100$ nm, refractive index $n = 3.5$, volume filling fraction $f = 38\%$) in air. Numerical simulation is performed with the spatio-temporal discretization corresponding to $\lambda_0/40$, which is twice finer than that used to produce Fig.~\ref{fig:n3.5}e in the main text. Pure exponential decay of $T(t)$ in time, over 12 orders of magnitude, is a hallmark of diffusive transport in the dielectric random system.
			\label{suppfig:Discretization_n3.5}}
		\end{center}
	\end{figure}

	(i) One key evidence for the absence of AL in 3D dielectric random media is the exponential decay of transmitted flux $T(t)$, shown in Fig.~1e of the main text. This result is obtained with the spatial grid size of $\lambda_0/20$. When the spatial grid is reduced to $\lambda_0/40$ and the temporal step size is also halved, the exponential decay of $T(t)$ persists over 12 orders of magnitude in Fig.~\ref{suppfig:Discretization_n3.5}, which confirms the diffusive transport.		
		
\begin{figure}[H]
		\begin{center}
			\includegraphics[width=\textwidth]{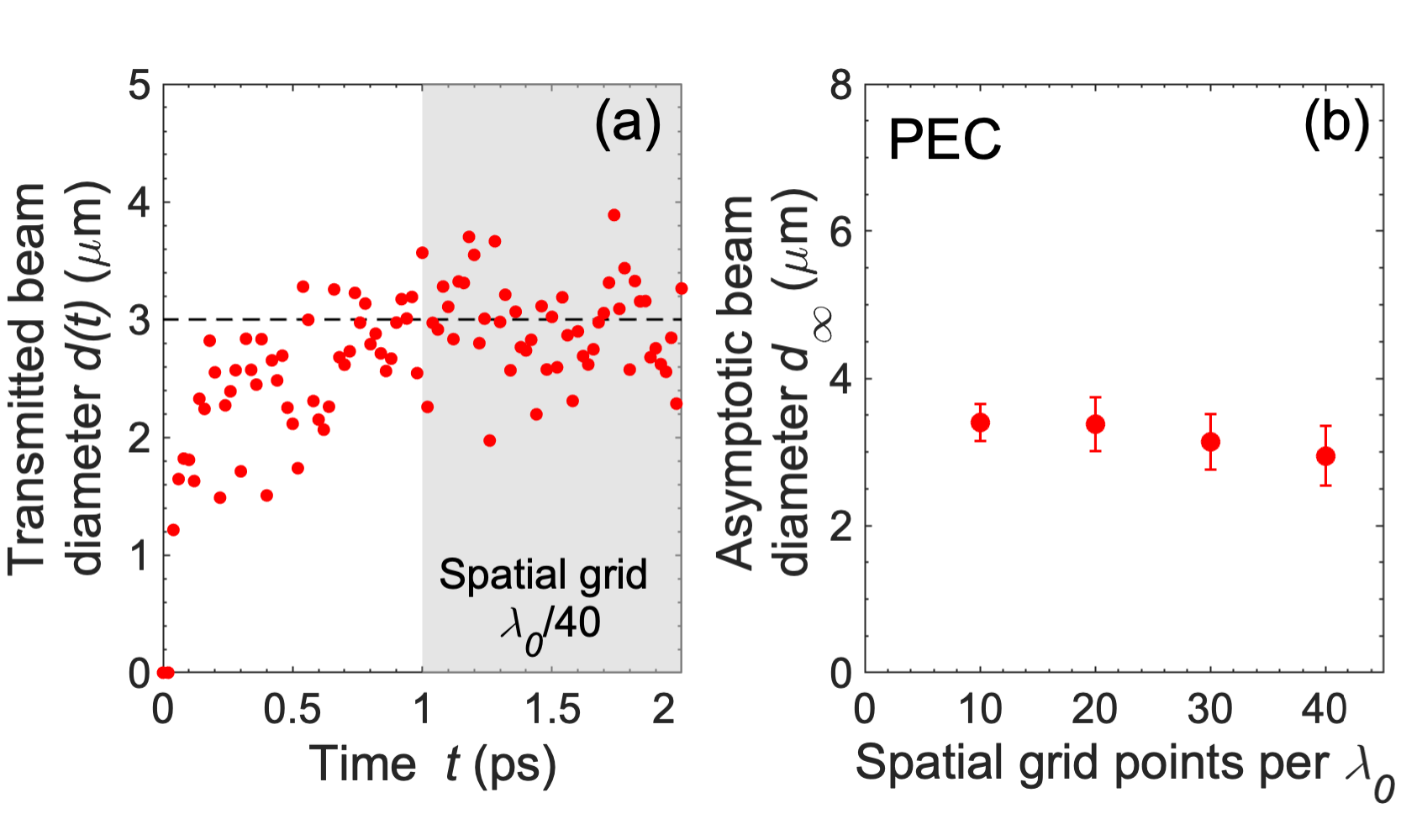}
			\caption{
			{\bf Robustness of the arrest of transverse spreading of transmitted beam in localized PEC systems against numerical discretization.}
			The simulation is schematically depicted in Fig.~\ref{fig:PEC_spreading}a of the main text, and the structure parameters (sphere radius $r = 50$ nm, volume filling fraction $f = 48\%$) are identical to those in Fig.~\ref{fig:PEC_spreading}d.
			a, Transverse diameter of the transmitted beam $d(t)$, obtained from numerical simulation (slab thickness $L = 1.3$ $\mu$m) with $\lambda_0/40$ resolution, exhibits a saturation consistent with the result with $\lambda_0/20$ resolution in Fig.~\ref{fig:PEC_spreading}d. 
			b, Arrest of the transverse spreading is observed in all simulations performed with numerical resolution from $\lambda_0/10$ to $\lambda_0/40$ with a consistent asymptotic transverse beam diameter $d_\infty$. Symbols and error-bars represent the mean value and standard deviation of $d(t)$ in the time interval $1$ ps $<t<2$ ps (gray area in panel a).
			\label{suppfig:Discretization_PEC}}
		\end{center}
	\end{figure}

	(ii) A tell-tale sign of AL in PEC composites is the arrest of transverse spreading of the transmitted beam, when an incident pulse is focused to a slab, as shown in Fig.~4c,d of the main text. This result is obtained with numerical resolution of $\lambda_0/20$. To test the robustness of this result, we vary the resolution from $\lambda_0/10$ to $\lambda_0/40$. Fig.~\ref{suppfig:Discretization_PEC}a shows the result with $\lambda_0/40$ resolution: the transverse diameter of transmitted beam $d(t)$ saturates to a constant value in time, consistent with the result of $\lambda_0/20$ resolution in Fig.~\ref{fig:PEC_spreading}d. The asymptotic beam diameter $d_\infty$ is plotted versus the numerical resolution in Fig.~\ref{suppfig:Discretization_PEC}b. It confirms the consistency of our numerical results: the arrest of transverse spreading persists in all simulations performed on the progressively finer meshes. Notably, the arrest is already seen in simulation with $\lambda_0/10$ resolution, despite a slight difference in the value of $d_\infty$.
		
	Since the Tidy3D implements the standard FDTD algorithm, the scaling of computing time with system size is the same as the standard FDTD method~\citesupp{2005_Taflove}. More specifically, the computing time scales as $N_x \, N_y \, N_z \, N_t$, where $N_x$, $N_y$, $N_z$ denote the number of spatial grid points in $x$, $y$, $z$ dimensions, $N_t$ is the total number of time steps. Typically the spatial grid size along $x$, $y$, $z$ is identical, and it is proportional to the time step.
		
	We note that the finite discretization introduces slight anisotropy and dispersion to the simulated system. To mitigate such effects, the spatio-temporal resolution could be further increased according to the system size, which would affect the scaling of computing time \citesupp{kruger2017solution}. However, this is not necessary in our simulations of random ensembles of dielectric or metal spheres, because the statistical properties like scattering and transport mean free paths are barely modified. By varying the spatio-temporal grid size, we find the small effects caused by finite discretization are simply equivalent to a tiny change of the refractive index $n$ of the medium and/or a tiny change of the volume filling fraction $f$. As shown in the manuscript, AL is absent in the dielectric systems for a broad range of $n$'s, whereas AL persists in a broad range of $f$'s in the PEC systems. 
		
	Our extensive tests of the numerical procedure and accuracy confirm the consistency and robustness of the results and conclusions in this manuscript. 
	
	\subsection{Spatial correlation function\label{suppsec:spatial_corr} }
	In order to avoid any residual spatial correlation of scattering structures, we adopt a uniform random distribution of scatterer centers. The structure factor obtained from Fourier transform of center positions is equal to unity~\citesupp{2018_Torquato_review}. The spheres with center spacing smaller than their diameter will overlap in space, and the overlapping region is assigned the refractive index equal to that of an isolated sphere.
	
\begin{figure}[H]
		\begin{center}
			\includegraphics[width=5.2in]{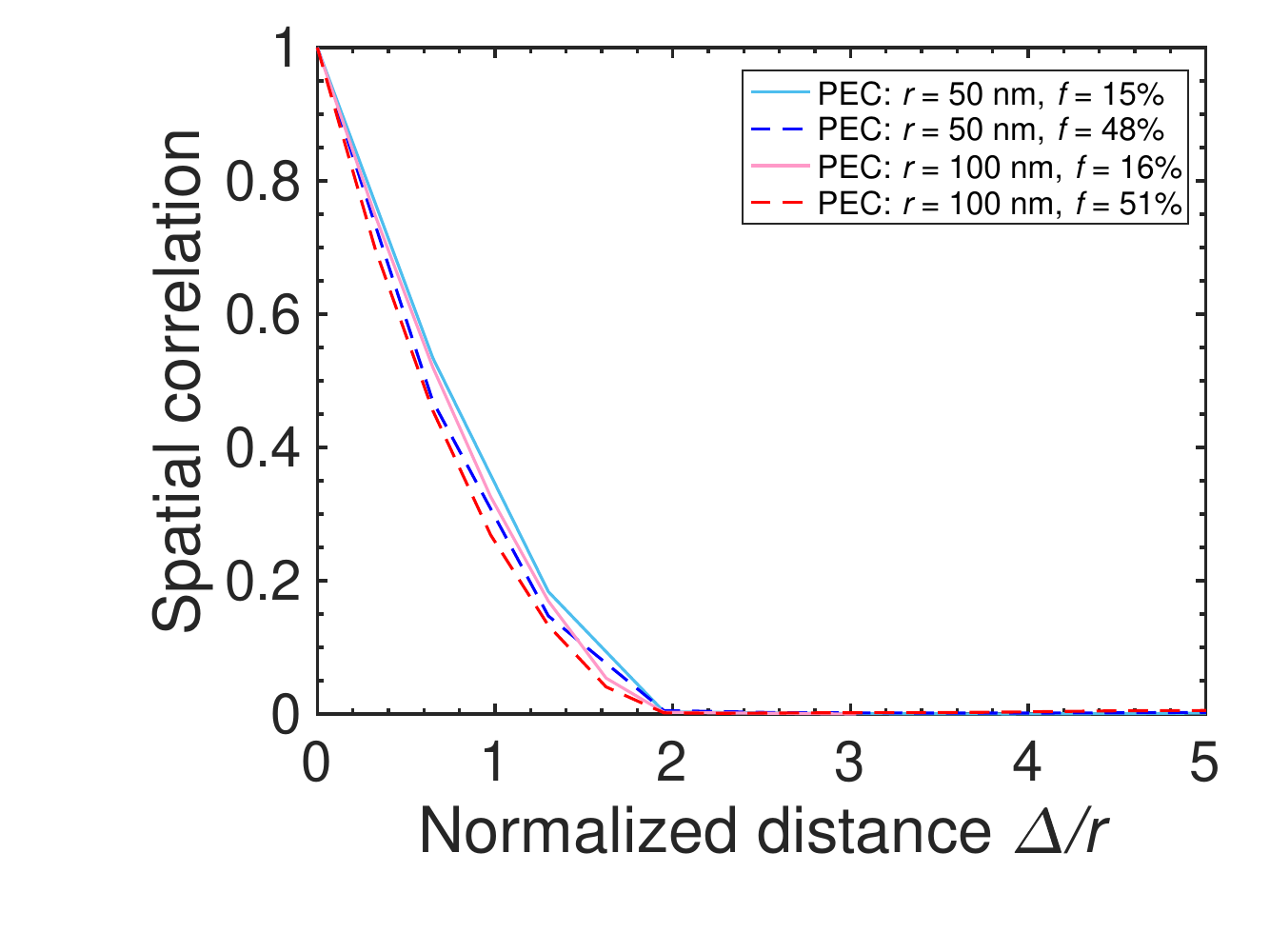}
			\caption{{\bf Spatial correlation in PEC composites.} Normalized spatial correlation function $C(\Delta)/C(0)$ 
			for random aggregates of overlapping spheres with radii $r=50$ nm, $100$ nm and volume filling fractions $f \simeq 15\%, 50\%$. Spatial correlation vanishes beyond one sphere diameter $2r$.
			\label{suppfig:spatial_correlation_function}}
		\end{center}
	\end{figure}

	We calculate the spatial correlation function $C(\Delta)=\langle h(\boldsymbol{\rho}) \, h(\boldsymbol{\rho}+{\bf \Delta})\rangle / \langle h(\boldsymbol{\rho})\rangle^2 - 1$, where, $\boldsymbol{\rho}$ denotes the spatial position, $h(\boldsymbol{\rho})$ is a binary function equal to $1$ inside the dielectric/PEC/metal scatterers and $0$ outside, and $\langle ...\rangle$ denotes averaging over $\boldsymbol{\rho}$, direction of ${\bf \Delta}$, and random ensemble. Fig.~\ref{suppfig:spatial_correlation_function} shows the normalized $C(\Delta)/C(0)$ with $h(\boldsymbol{\rho})$ obtained from the discretized structures in the actual FDTD calculations of systems with particle radii $r = 50$ nm and $100$ nm, and the PEC volume fractions of $f\sim 15\%$ and $\sim 50\%$. In all cases, the spatial correlation extends over the range $\Delta$ of one particle diameter $2 r$, as expected for a random arrangement of spherical particles.\\
	
	\subsection{Scattering mean free path}
	Scattering mean free path $\ell_s$ measures the average distance traveled between two consecutive scattering events. Its value in Figs.~\ref{fig:n3.5}c and \ref{fig:PEC_scaling}a is extracted from the attenuation of the coherent component of the incident field. We simulate systems with dimension $L_x =  100\lambda_0, L_y = L = 2\lambda_0$ for $n=3.5$ and PEC, $L_x =  100\lambda_0, L_y = L = \lambda_0$ in the $n=10$ case. The quantity $\langle E_x(x,y_0,z;\lambda)\rangle_x$ is computed by averaging over the long dimension (along $x$-axis) of the system for one particular cross section $y = y_0$. Average co-polarized field amplitude $|\langle E_x(x,y_0,z;\lambda)\rangle_x|$ decays exponentially at a rate $1/(2\ell_s)$, from which $\ell_s$ is obtained for each wavelength. The phase of $\langle E_x(x,y_0,z;\lambda)\rangle_x$ gives $k_{\text{eff}}$.
	
	Figures~\ref{suppfig:n3.5_ells},~\ref{suppfig:n10_ells},~\ref{suppfig:PEC_ells} show the amplitude and phase of coherent field $\langle E_x(x,y_0,z;\lambda)\rangle_x$, as well as its real and imaginary parts, in random aggregates of dielectric spheres and PEC spheres.  In the dielectric systems, spectral regions with strong attenuation, i.e. short scattering mean free path, are concentrated in the vicinity of Mie resonances of the constituent spherical particles. 	In contrast, in the PEC systems, scattering mean free path does not exhibit notable spectral features in Fig.~\ref{fig:PEC_scaling}a. In Figs.~\ref{suppfig:PEC_ells}a,e, this can be seen from a constant attenuation rate of the field amplitude. 
	
	\subsection{Transport mean free path}
	Transport mean free path $\ell_t$ corresponds to the average travel distance that is required to completely randomize the propagation direction. Transmittance of a continuous wave (CW) at wavelength $\lambda$ through a diffusive slab of thickness $L$ is  $T(\lambda)=(5/3)\ell_t(\lambda)/[L+2z_0(\lambda)]$, where $z_0(\lambda)$ is the extrapolation length~\cite{2007_Akkermans_book}. The value of $z_0(\lambda)$ can be estimated from the CW depth profile of internal intensity $I(z, \lambda)$ by linear extrapolation $I(z= L + z_0, \lambda) = 0$ (Fig.~\ref{suppfig:PEC_additional}c). Then the value of $\ell_t(\lambda)$ is extracted from $T(\lambda)$ with known $z_0(\lambda)$.

	Typically, $z_0$ depends on the refractive-index mismatch between the slab and the surrounding medium~\cite{2007_Akkermans_book}. However, our choice of the refractive index $n_{\text{eff}}$ for the surrounding layers eliminates this mismatch for a dielectric slab, leading to $z_0=(2/3)\ell_t$~\cite{2007_Akkermans_book}. Our numerical data suggest that this relation approximately holds for PEC composites embedded in air as well.
	
	\subsection{Diffusion coefficient}
	Diffusion coefficient $D=\ell_t v_E/3$ depends on both the transport mean free path $\ell_t$ and the energy velocity $v_E$~\citesupp{1996_Lagendijk}. The propagation of an optical pulse in a slab geometry makes it possible to directly extract $D$ from the decay of the transmittance,  $T(t) \propto\exp[-t/t_D]$, where $1/t_D=\pi^2D/(L+2z_0)^2$. In the localized slabs such as those in Fig.~\ref{fig:PEC}, the decay rate, $1/t_D$, changes with time. To obtain $D(t)$ in Fig.~\ref{fig:PEC}b, we use an exponential fit within a time window $[t-2\tau_D,t+2\tau_D]$, where $\tau_D$ is the arrival time of the peak of transmitted flux.
	
	\subsection{Scaling functions}
	To obtain the scaling function $d\log(T)/d\log(L)$ for the PEC slabs, we compute the logarithmic derivative of CW transmittance $T$ with respect to slab thickness $L$ for a range of values of $k \ell_s$. First, we use the wavelength dependence of $k \ell_s$ in Fig.~\ref{fig:PEC_scaling}a to map $k \ell_s$ to $\lambda$ for each $f$. Next, we compute $T(\lambda)$ for two systems with different thicknesses $L_1=2\lambda_0$ and $L_2=5\lambda_0$, where $\lambda_0 = 650$ nm denotes the center of the wavelength range of interest. Finally, we approximate $d\log(T)/d\log(L)$ by the finite difference $[\log(T_2)-\log(T_1)]/[\log(L_2)-\log(L_1)]$ for all $\lambda$ and all $f$. After eliminating $\lambda$, we obtain the dependence of $d\log(T)/d\log(L)$ on the Ioffe-Regel parameter $k \ell_s$ shown in Fig.~\ref{fig:PEC_scaling}c.
	
	\section{Absence of light localization in 3D dielectric systems with refractive-index contrast of $10$}
\begin{figure}[H]
		\begin{center}
			\includegraphics[width=\textwidth]{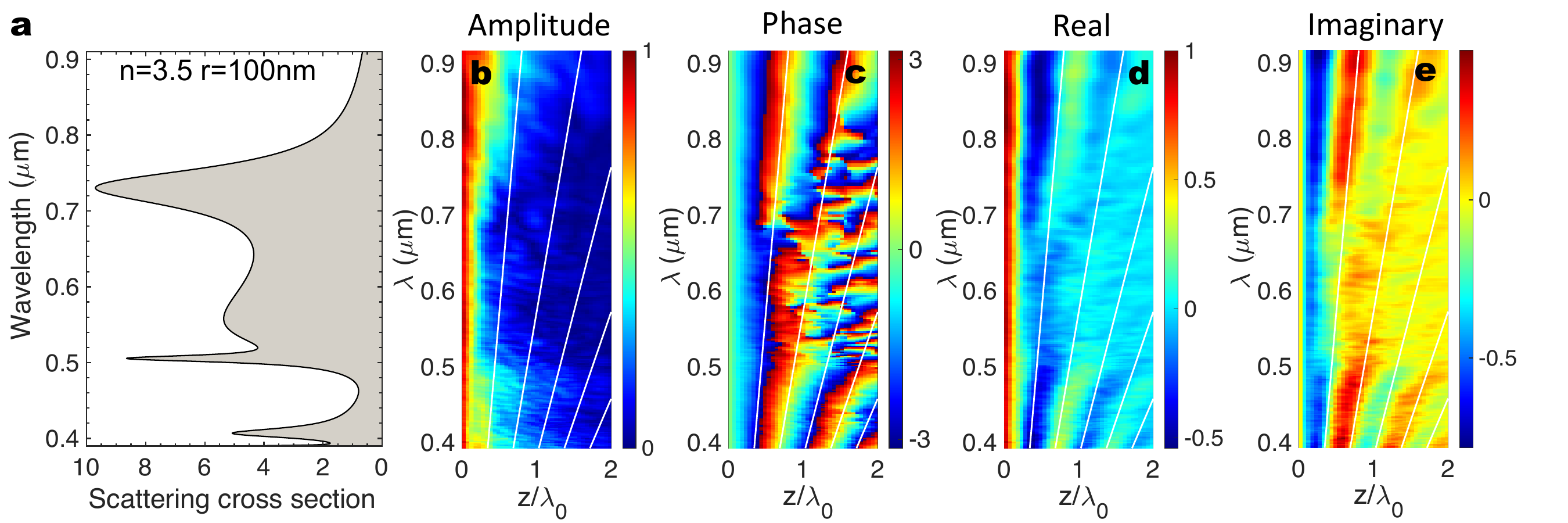}
			\caption{{\bf Extinction of coherent field in a dielectric slab of $n=3.5$ spheres.} 
				a, Scattering cross section of a single dielectric sphere of radius $r=100$ nm and refractive index $n=3.5$ in air. The spectral peaks correspond to Mie resonances.
				b,c,d,e, amplitude (b), phase (c), real (d) and imaginary (e) parts of coherent field $\langle E_x(x,y_0,z;\lambda)\rangle_x$ versus depth $z$ and wavelength $\lambda$ for dielectric filling fraction $f=29\%$. White lines represent $z=m\lambda_{\text{eff}}$, where $m$ is an integer and $\lambda_{\text{eff}}=\lambda_0/n_{\text{eff}}$. The coherent field amplitude experiences an enhanced extinction in the wavelength range 500--700 nm due to hybridized Mie resonances. 
				\label{suppfig:n3.5_ells}}
		\end{center}
	\end{figure}

\begin{figure}[H]
		\begin{center}
			\includegraphics[width=\textwidth]{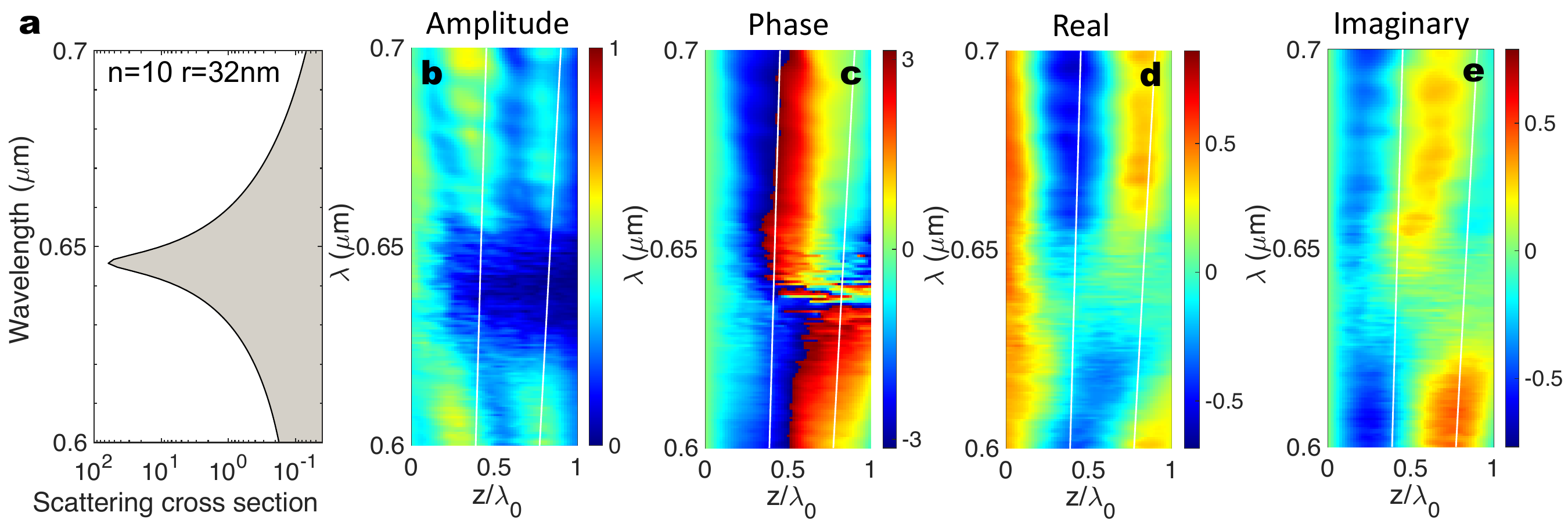}
			\caption{{\bf Extinction of coherent field in a dielectric slab of $n=10$ spheres.} 
				a, Scattering cross section of a single dielectric sphere of radius $r=32$ nm and refractive index $n=10$ in air. The peak represents the first Mie resonance. 
				b,c,d,e, amplitude (b), phase (c), real (d) and imaginary (e) parts of coherent field $\langle E_x(x,y_0,z;\lambda)\rangle_x$ versus depth $z$ and wavelength $\lambda$ for dielectric filling fraction $f=29\%$. White lines represent $z=m\lambda_{\text{eff}}$, where $m$ is an integer and $\lambda_{\text{eff}}=\lambda_0/n_{\text{eff}}$. The coherent field amplitude experiences an enhanced extinction in the spectral range 630--660 nm in the vicinity of the first Mie resonance. 
				\label{suppfig:n10_ells}}
		\end{center}
	\end{figure}

\begin{figure}[H]
		\begin{center}
			\includegraphics[width=5in]{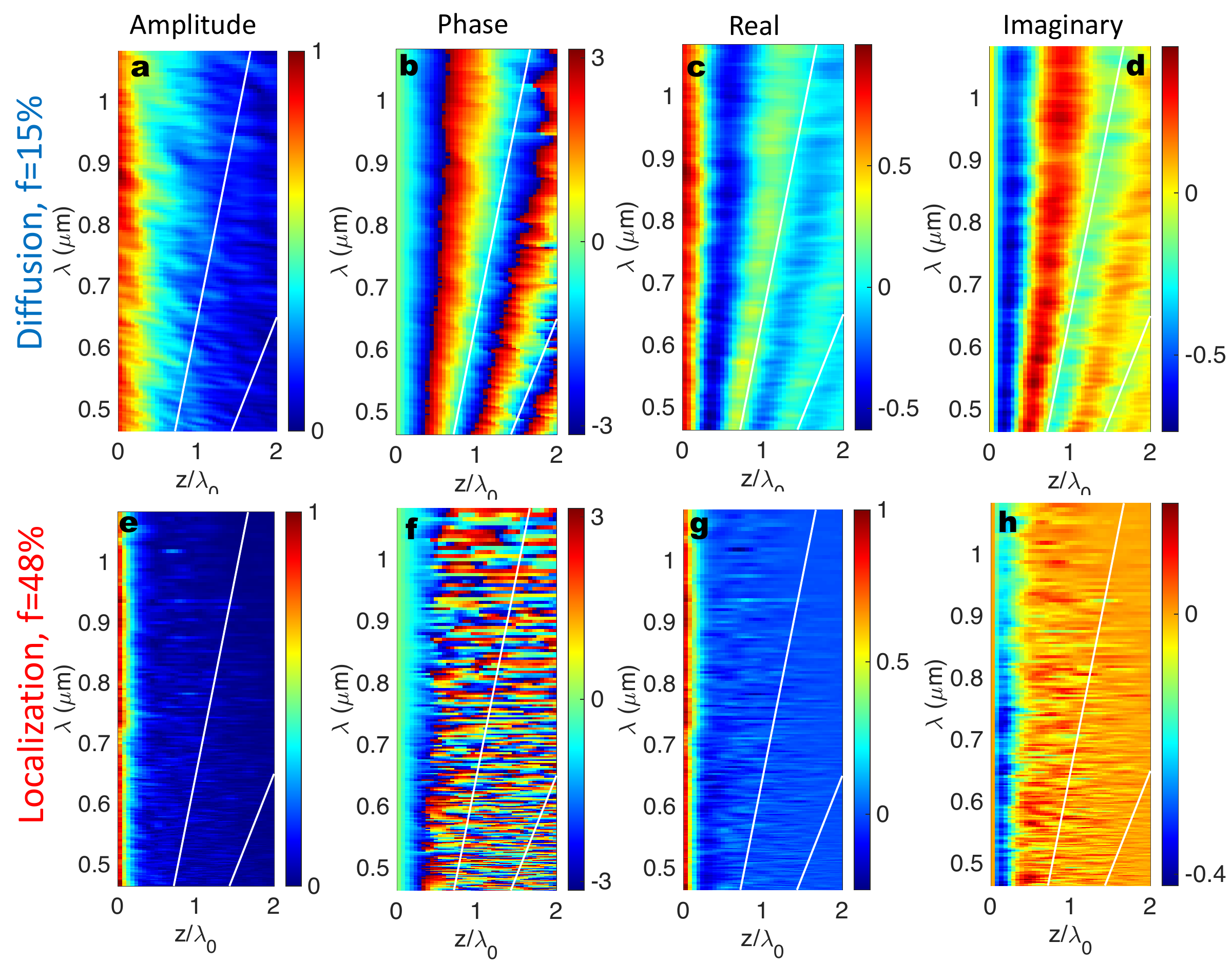}
			\caption{{\bf Comparison of coherent fields in diffusive and localized slabs of PEC spheres.}  
				Amplitude (a,e), phase (b,f), real (c,g) and imaginary (d,h) parts of coherent field $\langle E_x(x,y_0,z;\lambda)\rangle_x$ versus depth $z$ and wavelength $\lambda$ for PEC filling fraction $f=15\%$ (a,b,c,d) and $48\%$ (e,f,g,h). White lines represent $z=m\lambda$, where $m$ is an integer. In the localized slab ($f=48\%$), extinction of coherent field is much stronger and nearly independent of wavelength.
				\label{suppfig:PEC_ells}}
		\end{center}
	\end{figure}

\begin{figure}[H]
		\begin{center}
			\vskip -1cm 
			\includegraphics[width=5in]{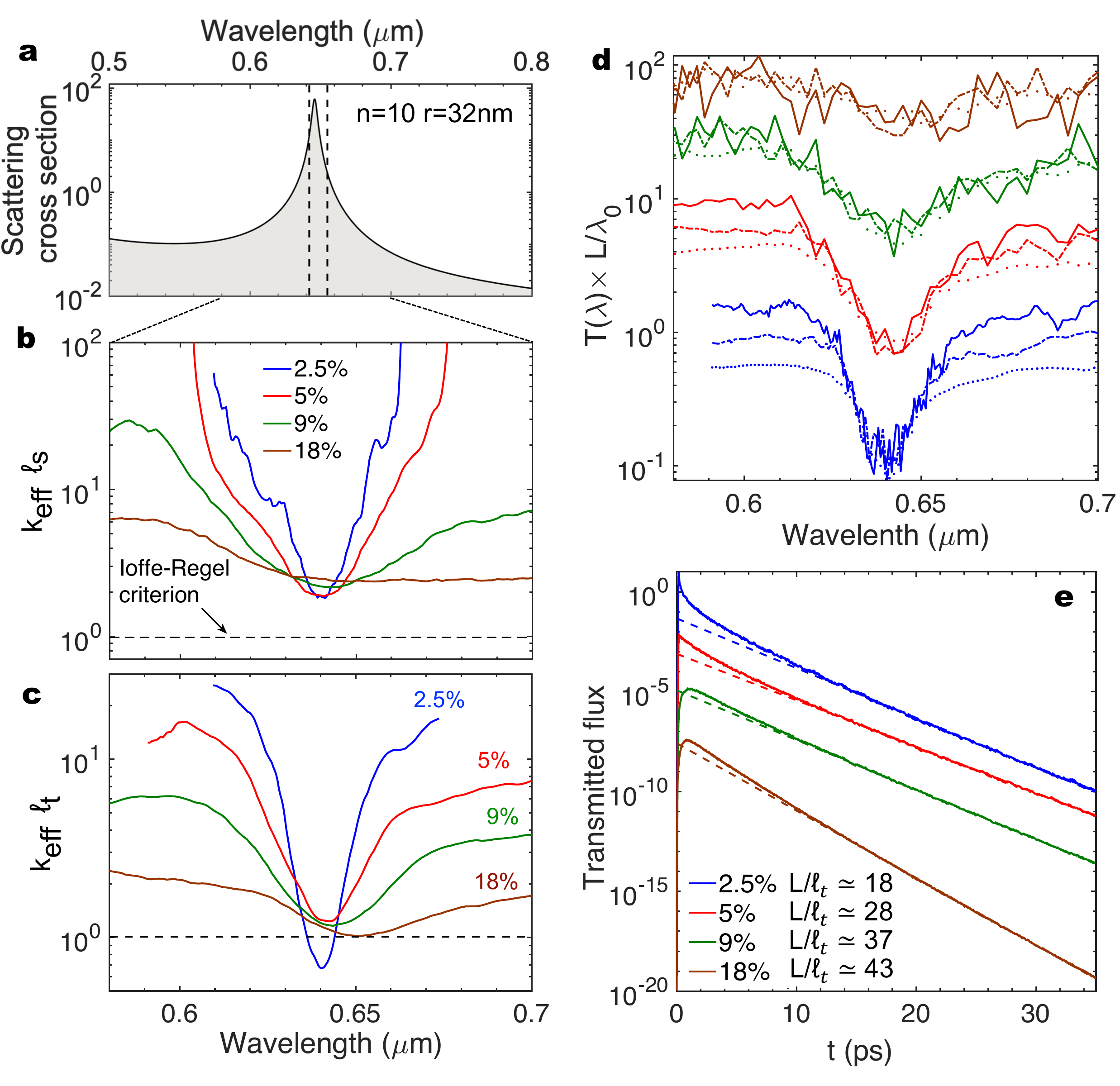}
			\caption{{\bf Resonant scattering and diffusion in random ensembles of $n=10$ dielectric spheres.} 
				a,~Scattering cross section of a single dielectric sphere of radius $r=32$ nm and refractive index $n=10$. The vertical dotted lines mark spectral width of the excitation pulse in panel e.
				b,~Resonantly enhanced scattering mean free path $\ell_s$ in the spectral vicinity of the first Mie resonance of single-particle scattering. The dielectric filling fraction $f = 2.5\%$ (blue), $f = 5\%$ (red), $9\%$ (green) and $18\%$ (brown). The minimum Ioffe-Regel parameter $k_{\text{eff}} \, \ell_s \sim 2$. The horizontal dashed line marks the Ioffe-Regel criterion $k_{\text{eff}} \, \ell_s =1 $  for 3D localization. 
				c,~Normalized transport mean free path $k_{\text{eff}} \, \ell_t$ in the spectral vicinity of the first Mie resonance, reflecting saturation by dependent scattering between $f = 2.5\%$ and $18\%$.  
				d, CW transmittance $T(\lambda)$ for three slab thicknesses $L/\lambda_0=1,2,4$ (solid, dashed, and dotted lines). $T(\lambda)$ multiplied by $L/\lambda_0$ remains nearly independent of $L$ at wavelengths around minimum transmission, reflecting the diffusive scaling of $T \propto 1/L$.  Results are color-coded as in panels b,c. For $f=5\%$, 9\% and 18\%, the curves are shifted up vertically by one, two and three decades for legibility.  	
				e,~Transmittance of 3D slabs with thickness $L=2\times\lambda_0$ for pulsed excitation, showing an exponential decay (dashed lines) in time at long delay. The decay rate is determined by the minimum diffusion coefficient within the pulse bandwidth. Legend shows values of $L/\ell_t $ in each system. For $f=5\%$, 9\% and 18\%, the curves are shifted down vertically by two, four and six decades for legibility.
				\label{suppfig:n10}}
		\end{center}
	\end{figure}

	This section reports the results of numerical simulation for a 3D slab of dielectric spheres with $n=10$. For ease of comparison, we adjust the sphere radius to $r=32$ nm, so that the first Mie resonance occurs at a wavelength similar to that for $n=3.5$ sphere (compare Figs~\ref{suppfig:n3.5_ells}a and \ref{suppfig:n10_ells}a). Scattering cross section of a single sphere of $n=10$ is exceedingly large (Fig. \ref{suppfig:n10}a), leading to strong dependent scattering already at small volume filling fractions, and limiting Ioffe-Regel parameter to $k_{\text{eff}} \, \ell_s \gtrsim 1$ in Fig. \ref{suppfig:n10}b. The minimal value of $k_{\text{eff}} \, \ell_s$ is almost identical for $f=2.5\%,\,5\%,\,9\%$, and starts to increase for $f=18\%$ due to dependent scattering. The dependent scattering also results in less variation of the normalized transport mean free path $k_{\text{eff}} \, \ell_t$ with $\lambda$ in Fig.~\ref{suppfig:n10}c, similar to the scattering mean free path in Fig.~\ref{suppfig:n10}b.
	
	Diffusive nature of wave propagation in $n=10$ dielectric systems can be seen from scaling of the CW transmittance with slab thickness $L$. In Fig.~\ref{suppfig:n10}d we plot transmittance multiplied by $L$ in the $f=2.5\%,\,5\%,\,9\%,\,18\%$ slabs with three different values of $L/\lambda_0=1,2,4$. The largest slab thickness $L/\lambda_0=4$ corresponds to $L/\ell_t>36\gg1$ for all $f$. The overlap between the curves with different $L$ in the spectral range of the first Mie resonance is a direct manifestation of the diffusive scaling $T\propto 1/L$. 
	
	Furthermore, we numerically calculate dynamic transmittance under pulsed excitation, $T(t)$. For thick slabs ($L/\ell_t\gg 1$), it does not exhibit deviation from the diffusive transport: at $t\gg \tau_D$, the decay is still exponential (Fig. \ref{suppfig:n10}e). 
	
	The above results confirm light transport is diffusive in 3D dielectric random systems with $n=10$ spheres. 
	
	\section{Difference between PEC and dielectric scattering}
	
\begin{figure}[H]
		\begin{center}
			\includegraphics[width=5in]{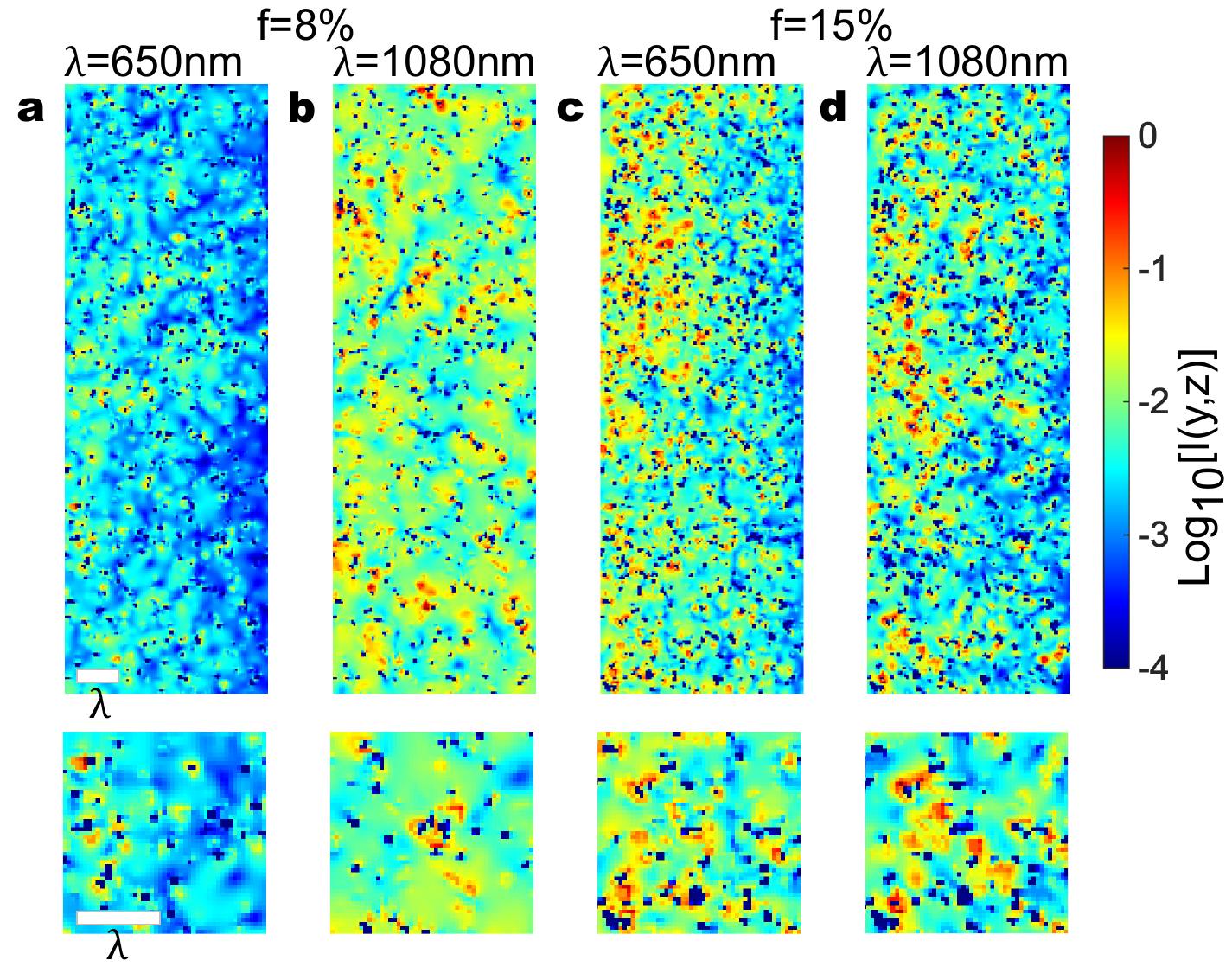}
			\caption{{\bf Spatial intensity distribution inside disordered PEC system.} 
				Field intensity at a $(y,z)$ cross-section inside a random ensemble of PEC spheres with radius $r=50$ nm. PEC filling fraction is $f=8\%$ in panels a,b and $f=15\%$ in panels c,d. $\lambda = 650$ nm in panels a,c and 1080 nm in panels b,d. Bottom row shows a magnified view of each distribution in the top row, revealing strong intensity enhancement in air voids between PEC particles due to coupled resonances. The size of the voids can be much smaller than wavelength $\lambda$ indicated by the white scale bar.
				\label{suppfig:PEC_red_shift}}
		\end{center}
	\end{figure}

	Besides the absence of longitudinal fields in PEC composites, additional differences with respect to dielectric media facilitate Anderson localization in PEC:

	\begin{enumerate}
	\item Isolated PEC spheres have strong backscattering~\cite{1981_van_de_Hulst}. It results in a highly anisotropic angular scattering pattern, making $\ell_t < \ell_s$. This is very different from the dielectric spheres. 

	\item Collective scattering resonances are created by two or more adjacent PEC spheres. The spatial arrangement of scatterers in this work lacks spatial correlations in sphere positions, creating a random distribution of distance between them. Even at the PEC volume fraction of $f = 8\%$, the gap in-between two PEC spheres can be much narrower than $\lambda = 650$ nm, and the field intensity in the vicinity of the gap is strongly enhanced, as seen in Figs.~\ref{suppfig:PEC_red_shift}a. Such an enhancement persists at $\lambda = 1080$ nm in Figs.~\ref{suppfig:PEC_red_shift}b, despite the wavelength is much greater than the particle radius $r$ = 50 nm, leading to a small size parameter $k r\simeq 0.3<1$. Already for volume fraction of $f=8\%$, a sufficiently broad distribution of gap sizes creates a large number of red-shifted resonances, enhancing the overall scattering at longer wavelengths in Fig.~\ref{fig:PEC_scaling}a, and making $\ell_s$ well below the prediction~\citesupp{1999_van_Rossum} of the independent scattering approximation (ISA) $\ell_s=[\rho_{sca}(f)\,\sigma_{sca}(\lambda)]^{-1}$, where $\rho_{sca}(f)$ is the number density of scatterers and $\sigma_{sca}(\lambda)$ is the scattering cross section of a single PEC sphere. With an increase of $f$ to $15\%$ in Figs.~\ref{suppfig:PEC_red_shift}c,d, additional voids formed by three or more PEC spheres create even a larger variety of scattering resonances at different wavelengths rendering $\ell_s$ nearly constant with~$\lambda$.

	\item At high volume fraction $f$ of PEC scatterers, light propagation through air voids is suppressed. In contrast to a dielectric composite which becomes transparent when the dielectric volume fraction approaches 100\%, a PEC composite with $f$ approaching 100\% becomes a perfect mirror, as air voids no longer percolate through the system. In our simulations, Anderson localization takes place at PEC volume fractions well below the air percolation threshold (96.6\%) \citesupp{1981_Percolation_of_spheres,1984_Elam_Percolation_of_spheres}. The air voids are still connected by narrow channels through which light may propagate. However, when the typical size of air voids between the PEC scatterers is on the order of the wavelength, light can no longer `squeeze' through the narrow channels. Thus AL occurs at the wavelength comparable to or larger than the typical size $a$ of air voids between PEC scatterers.
	
	For a given PEC volume fraction $f$, air void size $a$ is proportional to PEC sphere radius $r$. A random system with smaller spheres have a larger number of voids, but the voids themselves have smaller sizes, since the total volume of all voids is fixed by $f$. The smaller voids make it harder for light to propagate through a system with smaller spheres, and localization will take place at lower $f$. This prediction agrees to our numerical results of lower critical $f$ for AL with $r = 50$ nm PEC spheres than $r = 100$ nm spheres. The above argument is for fixed wavelength $\lambda$. In a PEC system with given $r$ and $f$, the typical void size $a$ is fixed, increasing $\lambda$ will make it more difficult for light to `squeeze' through voids, facilitating AL at long wavelengths. Consequently, for a given $r$, the localization condition of $a \sim \lambda$ is fulfilled at a smaller $f$ at longer $\lambda$, which agrees to our numerical results.
	\end{enumerate}
	
	\section{Additional evidence for Anderson localization in PEC system}
\begin{figure}[H]
		\begin{center}
			\includegraphics[width=\textwidth]{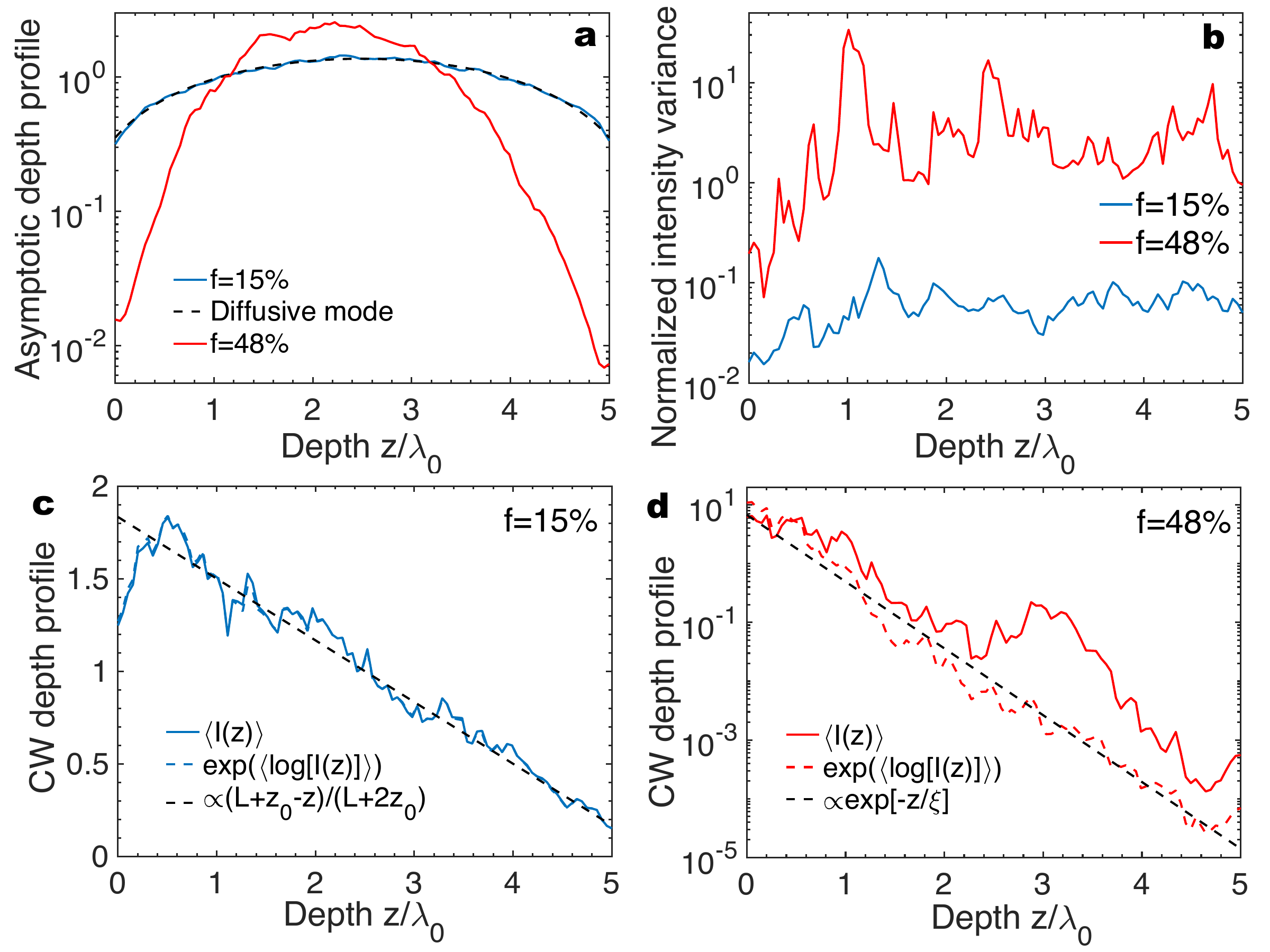}
			\caption{{\bf Additional evidence for Anderson localization in disordered PEC.} 
				a, Depth profile $\langle I(x,y,z;t\rightarrow\infty)\rangle_{x,y}$ inside the PEC slab of $f = 15\%$ (blue solid line) and $48\%$ (red solid line). The lowest-order diffusive mode profile (black dashed line) matches that of $f =15\%$ (blue).  
				b, Normalized variance of cross-section integrated intensity, showing two orders of magnitude enhancement of intensity fluctuations for $f = 48\%$ over $f = 15\%$. 
				c,d, CW intensity depth profiles $\langle I(z,\lambda)\rangle_{\lambda}$ (solid lines) and $\exp[\langle\log[I(z,\lambda)]\rangle_{\lambda}]$ (dashed lines), compared to linear decay for $f = 15\%$ and exponential decay for $f = 48\%$ (black dashed lines).  
				\label{suppfig:PEC_additional}}
		\end{center}
	\end{figure}

	We have examined the intensity distributions inside 3D slabs of PEC spheres with $r=50$ nm in dynamical simulations. Fig.~\ref{suppfig:PEC_additional}a shows the cross-section averaged depth profiles at very long delay times, $\langle I(x,y,z;t\rightarrow\infty)\rangle_{x,y}$, for PEC filling fractions of $15\%$ and $48\%$. In the diffusive slab of $f = 15\%$, the asymptotic depth profile matches the first eigenmode of the diffusion equation, $\sin[\pi(z+z_0)/(L+2 z_0)]$. In contrast, the cross-section averaged intensity inside the localized slab with $f = 48\%$ exhibits much larger variation with depth $z$. The faster decay towards the surfaces reflects stronger confinement of energy near the center of the slab. 
	
	Fluctuation of intensity or transmission coefficients is a powerful criterion of localization~\cite{chabanov00}. Fig.~\ref{suppfig:PEC_additional}b depicts spectral fluctuation of the cross-section average intensity $I(z,\lambda)=\langle I(x,y,z;\lambda)\rangle_{x,y}$, within the wavelength range of 600--700 nm. The normalized variance \\ ${\rm var}_\lambda[I(z,\lambda)] / \langle I(z,\lambda)\rangle_\lambda^2$ corresponds to the magnitude of long-range correlation $C_2$ \citesupp{2014_Sarma_Cofz}. In the PEC slab of $f=15\%$,  $C_2 \ll 1$, typical of a diffusive system. When $f$ increases to $48\%$, $C_2 \geq 1$, as expected for localized systems\citesupp{1999_van_Rossum,2004_Yamilov_passive_corr},\cite{chabanov00}.
	
	Inside a diffusive system, the CW intensity decays linearly with the depth~\cite{2007_Akkermans_book}, and exhibits low fluctuations from realization to realization. This is confirmed in Fig.~\ref{suppfig:PEC_additional}c by comparing $\langle I(z,\lambda)\rangle_{\lambda}$ and $\exp[\langle\log[I(z,\lambda)]\rangle_{\lambda}]$, which indeed agree well in the PEC slab of $f= 15\%$. In contrast, in the PEC slab of $f= 48\%$, these two quantities are markedly different owning to the fact that strong intensity fluctuations lead to log-normal distribution for localized systems \citesupp{2000_Mirlin}. The depth profile exhibits a roughly exponential decay, as a result of AL. 
	
\begin{figure}[H]
		\begin{center}
			\includegraphics[width=2.75in]{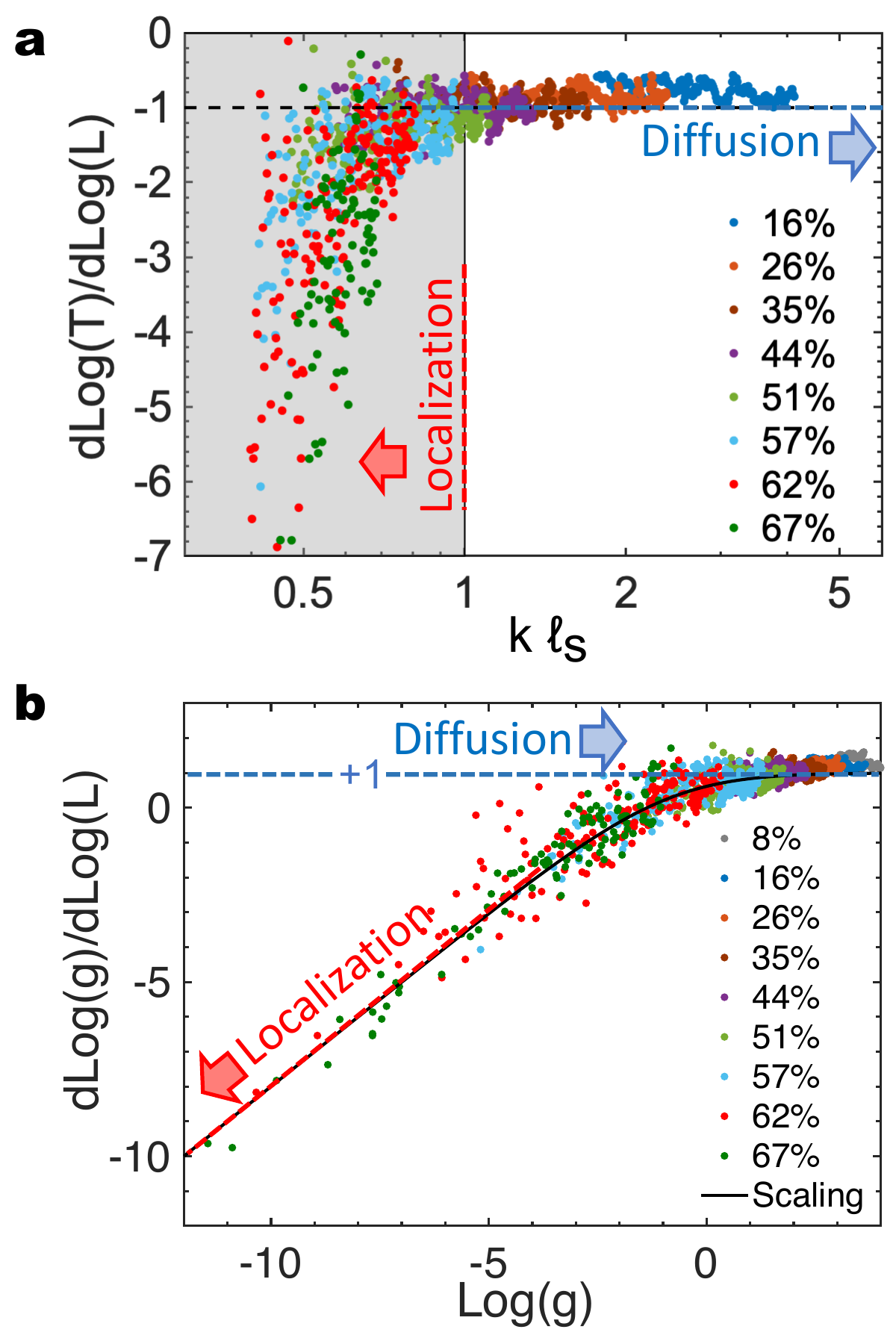}
			\caption{{\bf Diffusion-localization transition in another PEC system with larger spheres.} The scaling behavior reported in Fig.~\ref{fig:PEC_scaling}b,c is reproduced for a system of larger PEC spheres with radius $r=100$ nm, despite the diffusion-localization transition occurs at a different volume filling fraction $f$ of PEC spheres. In a, blue dashed line denotes diffusive scaling $T\propto L^{-1}$; red dashed line marks $k \ell_s = 1$. In b, blue and red dashed lines denote diffusive and localized scaling $g\propto L$ and $g\propto\exp(-L/\xi)$ respectively, where $\xi$ is the localization length.
				\label{suppfig:PEC_r100_scaling}}
		\end{center}
	\end{figure}

	To further confirm AL in PEC slabs, we repeat the calculations reported in Fig.~\ref{fig:PEC_scaling}c,d for another system with larger PEC spheres ($r = 100$ nm). Figure~\ref{suppfig:PEC_r100_scaling} demonstrates that transmittance and conductance exhibit similar scaling behaviors as for the PEC slabs of $r$ = 50 nm in Fig.~\ref{fig:PEC_scaling}c,d. Although the diffusion-localization transition occurs at a different volume filling fraction $f$ of PEC spheres, the scaling behaviors with respect to the Ioffe-Regel parameter $k \, \ell_s$ and to the dimensionless conductance $g$ are reproduced.
	
	\section{Anderson localization in real metals}
	To demonstrate the possibility of AL in realistic metallic systems, we simulate real metals using parameters reported in literature.
	The dielectric constant of a metal is related to the conductivity  $\sigma(\omega)$ via $\epsilon(\omega)=1+i \sigma(\omega)/\omega\epsilon_0$, where $\epsilon_0$ is electric permittivity of vacuum. In the Drude model, the conductivity of a metal is given by $\sigma(\omega)=\sigma_0/(1-i\omega\tau)$, where $\sigma_0$ is the static conductivity and $\tau$ is the relaxation time. 
	
	\subsection{Microwave regime}
	At microwave frequencies, $\omega \ll 1/\tau$ and $\epsilon(\omega)\simeq 4\pi\sigma_0 i/\omega$. The penetration depth of electric field into the metal is characterized by the skin depth  $\sim (\lambda_0/2\pi)\, (\omega/2\pi\sigma_0)^{1/2}$. Common metals like silver, aluminum, and copper have low loss and large conductivity at microwave frequencies, and their skin depth is much shorter than the wavelength. For example, at $20$ GHz where the wavelength is $1.5$ cm, the skin depth is less than $1\ \mu$m~\citesupp{2017_Yi_Plasmonics_book}.
	
	To simulate microwave transport in 3D aggregates of metallic spheres, we generate random arrangements of overlapping spheres using the same procedure as for the PEC systems. The sphere diameter of $0.56$ cm is four orders of magnitude larger than the skin depth of crystalline metals like silver, aluminum, copper. Thus the microwave barely penetrates into the metallic scatterers, and the absorption loss is negligible. When we use the conductivity $\sigma_0=3.8\times 10^{7}$ ${\rm \Omega^{-1}/m}$ of crystalline aluminum reported in literature~\citesupp{2017_Yi_Plasmonics_book}, the simulation results barely deviate from those for PEC.
	
\begin{figure}[H]
		\begin{center}
			\includegraphics[width=4in]{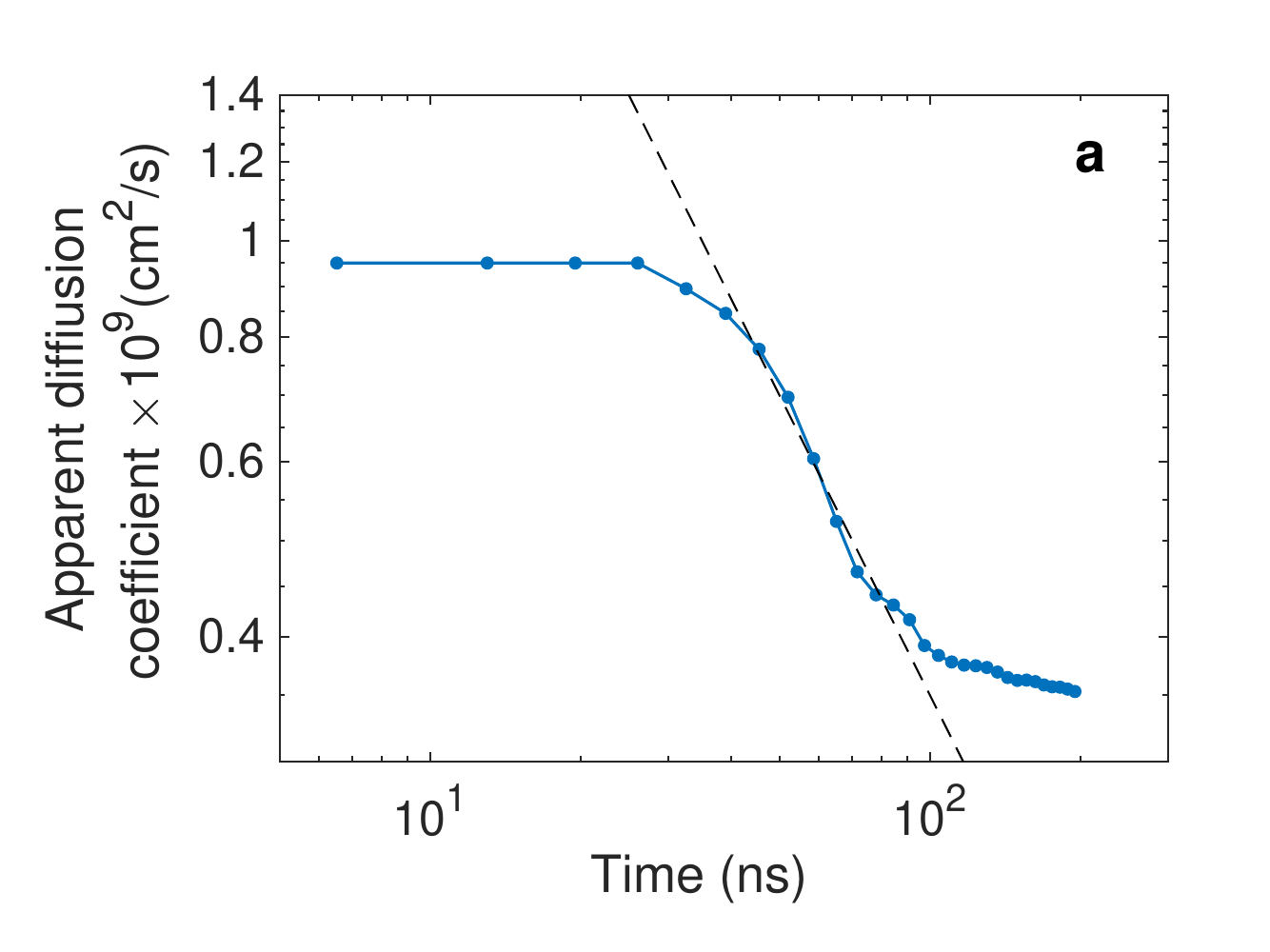}
			\includegraphics[width=4in]{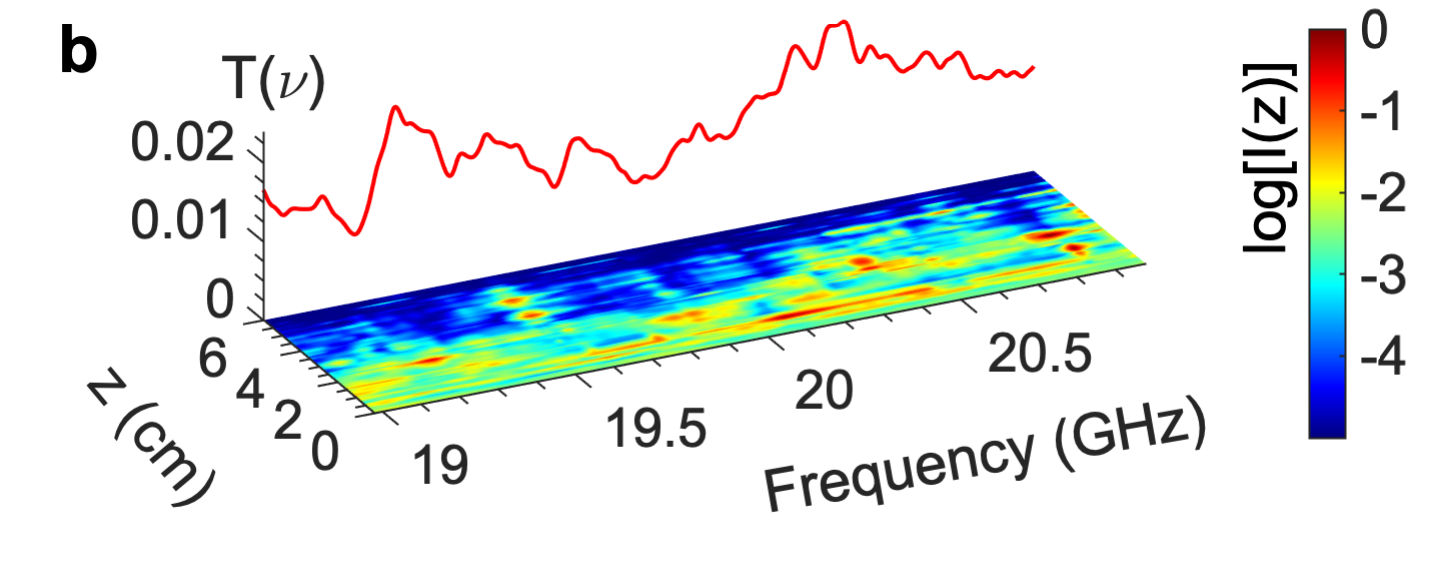}
			\caption{{\bf Localization of microwaves in a 3D real-metal composite.} A random aggregate of aluminum spheres with radius $r = 0.28$ cm, realistic conductivity $\sigma_0=3.8\times 10^{4}$ ${\rm \Omega^{-1}/m}$, and volume filling fraction $f = 60\%$ localizes microwave of frequency around $20$ GHz in a slab of thickness $L$ = 6 cm. 
			a, Time-resolved apparent diffusion coefficient $D(t)$, extracted from the temporal decay rate of $T(t)$, decreases as $1/t$, due to Anderson localization. Eventually it saturates to the value set by metal absorption. 
			b, Wavelength-resolved transmittance $T(\lambda)$ (red line) exhibits fluctuations. Color map: depth profile of average intensity $\langle I(x,y_0,z;\lambda)\rangle_x$ inside the slab at different wavelengths, highlighting spatially localized and necklace-like states.
			\label{suppfig:microwave_Doft}}
		\end{center}
	\end{figure}

	However, experimentally fabricated metallic structures have additional loss due to polycrystallinity, surface defects, oxide layers, etc. To take these into account, we lower the conductivity $\sigma_0$ so that the simulated transport properties match the experimental values in Ref.~\cite{genack91}. Specifically, we simulate dynamic transmittance in a diffusive slab of aluminum spheres with same volume fraction $f=35\%$ and slab thickness $L$ = 6 cm as in the experiment. When the conductivity is reduced to $\sigma_0=3.8\times 10^{4}$ ${\rm \Omega^{-1}/m}$, our simulated diffusion coefficient $D=1.9\times 10^9$ cm$^2$/s and absorption coefficient $\alpha=(D\tau_a)^{-1}\simeq 0.2$ cm$^{-1}$ both agree to the experimental values at 20 GHz. Using these realistic parameters, we simulate 3D metallic composites with high volume fraction of $f=60\%$.  As shown in Fig.~\ref{suppfig:microwave_Doft}a, the apparent diffusion coefficient, obtained from the temporal decay of the transmittance without taking absorption into account, decreases in time as $1/t$, similarly to its behavior in PEC, until it reaches the value set by the absorption. The simulated field intensity distribution inside the system, in Fig.~\ref{suppfig:microwave_Doft}b, provides evidence of spatial confinement of microwave, similar to the result for PEC in Fig.~\ref{fig:PEC}e. The most conclusive evidence of AL is the arrest of transverse spreading of transmitted wave field. It is insensitive to absorption and is shown in the main text in Fig.~\ref{fig:PEC_spreading}e.
	
	We believe the tell-tale experimental sign of AL is the arrest of transverse spreading of transmitted wave with a focused incident pulse. The potential pitfalls are background signals or possible emission by the metal upon microwave excitation. However, these signals are typically broadband and incoherent with the incident wave. If the incident microwave has a narrow spectral band, the transmitted field may be measured coherently, e.g., using the homodyne technique common for microwaves, and those background and incoherent signals will be discarded. Hence, we propose an experiment with a frequency-tunable narrowband microwave source: focus the incident microwave to the front surface of a slab and scan the frequency, perform an interferometric measurement of transmitted field distribution near the slab back surface at each frequency (using a local oscillator), finally Fourier transform the spectral fields to reconstruct the temporal evolution of transmitted field for an incident short pulse in order to observe the arrest of transverse spreading.
	
	\subsection{Optical regime}
\begin{figure}[H]
		\begin{center}
			\includegraphics[width=4in]{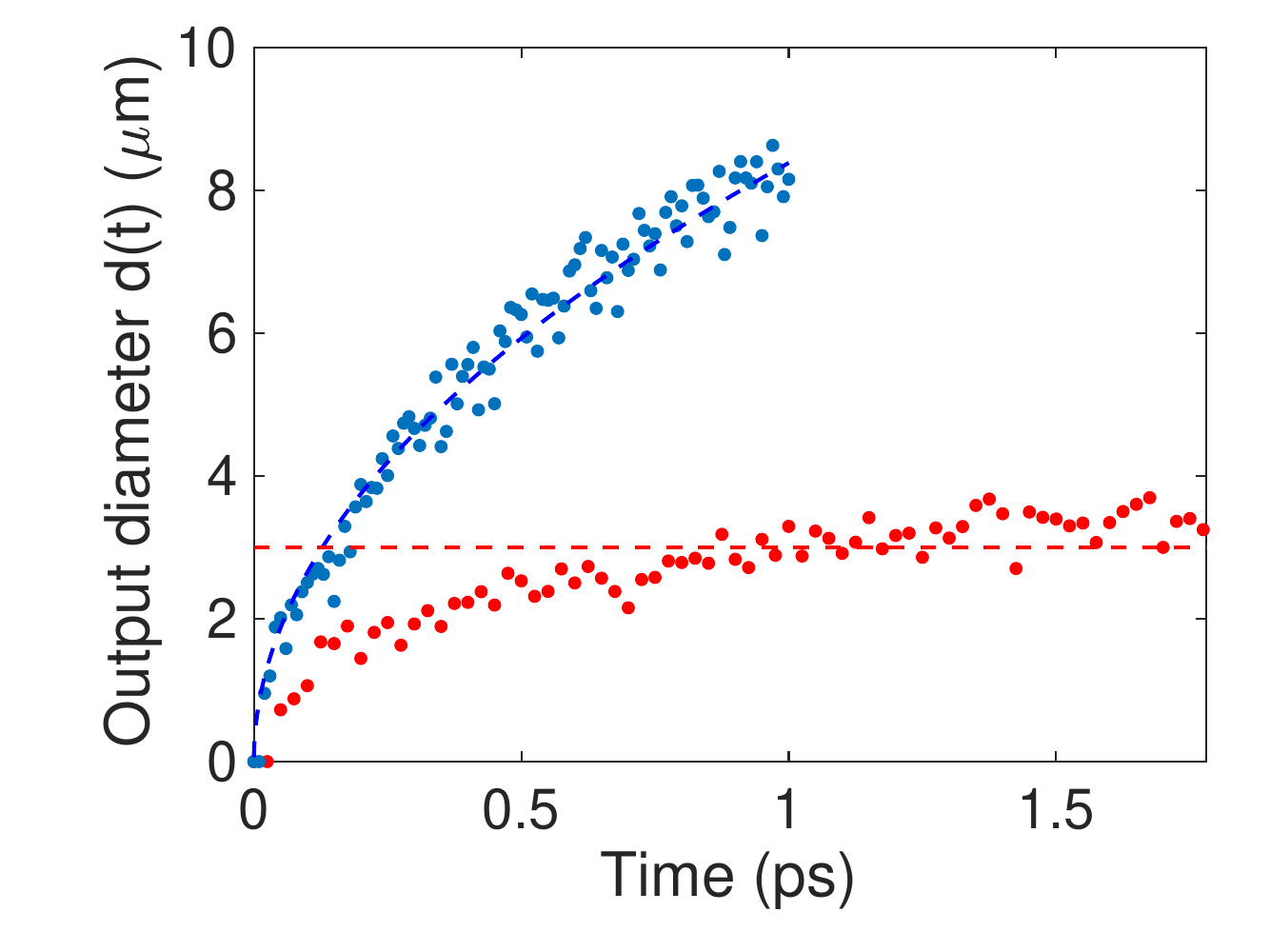}
			\caption{{\bf Arrest of transverse spreading of visible light in 3D random aggregates of silver particles.} Transverse diameter of the transmitted light $d(t)$, with a tightly-focused incident pulse of center wavelength $\lambda_0 = 650$ nm, increases as $\sqrt{t}$ in the diffusive slab ($f=15\%$, blue dots), but it saturates to a constant value in the localized slab ($f=48\%$, red dots). The slab thickness $L$ = 2 $\mu$m. Silver spheres have radius $r$ = 50 nm, $\sigma_0\simeq 6.1\times 10^7$ ${\rm \Omega^{-1}/m}$ and $\tau \simeq 3.7\times 10^{-14}$ s.
			\label{suppfig:optical_spreading}}
		\end{center}
	\end{figure}

	Here we investigate the possibility of AL of visible light in 3D metallic nanostructures. Even for low-loss metals like gold and silver, the skin depth is $\sim 25$ nm at $\lambda=650$ nm~\citesupp{2017_Yi_Plasmonics_book}, comparable to the nanoparticle diameter of 100--200 nm. A significant penetration of light field inside the metal particles makes them deviate from the PEC, which expels the fields completely. Another consequence of the penetration is a notable loss. We simulate silver, which is widely used in nano-plasmonics and meta-materials, using realistic parameters reported in the literature. We adopt the Drude model of $\epsilon(\omega)$ with $\sigma_0\simeq 6.1\times 10^7$ ${\rm \Omega^{-1}/m}$ and $\tau \simeq 3.7\times 10^{-14}$~s from Ref.~\citesupp{1985_Ordal_Drude_model_parameters}. Despite absorption and deviation from PEC, we still observe the arrest of transverse spreading in Fig.~\ref{suppfig:optical_spreading}. 
	
	Compared to 2D nanostructures, 3D nanoporous metals have a much larger (internal) surface area, leading to a wide range of applications in photo-catalysis \citesupp{linic2011plasmonic}, optical sensing \citesupp{chen2009nanoporous}, energy conversion and storage \citesupp{mascaretti2019plasmon}. Metallic nanostructures have been widely explored to enhance Raman scattering, second-harmonic generation, etc., because they can produce `hot spots' (giant local fields) to boost optical nonlinearities. Also metallic nanoparticles have been used for random lasing, as their strong scattering of light improves optical feedback \citesupp{wang2017nanolasers, gomes2021recent}. Our simulation results suggest the possibility of AL in 3D metallic nanostructures at optical frequencies. Light localization in such structures will have a significant impact on optical nonlinear, lasing, photochemical processes and related applications.
	
	\newpage
	\bibliographystylesupp{Nature}
    \bibliographysupp{supp.bbl}	
	\newpage
	\newpage
	\newpage
	\newpage
	\newpage
	\newpage
	\newpage
	\newpage
	\newpage
	\newpage
	
	\newpage
	\newpage
\end{document}